\documentclass[pra,twocolumn,showpacs,groupaddress,floatfix]{revtex4}
\usepackage{graphicx}
\usepackage[dvips]{color}
\usepackage{amsmath,amssymb,bm}
\usepackage{times}
\usepackage{txfonts}

\newcommand{\Rb}{$^{85}$Rb}
\newcommand{\Rbdimer}{\Rb$_{\,2}$}
\newcommand{\Rbtrimer}{\Rb$_{\,3}$}
\newcommand{\Eins}{\mathbf{1}}

\renewcommand{\vec}[1]{\bm{\mathrm{#1}}}
\newcommand{\ket}[1]{| #1\rangle} \newcommand{\bra}[1]{\langle #1|}
\newcommand{\braket}[2]{\langle #1|#2\rangle}

\newcommand{\psidim}{\psi_\mathrm{b}}
\newcommand{\Edim}{E_{\mathrm{2B}}}
\newcommand{\Etri}{E_{\mathrm{3B}}}

\newcommand{\rel}{\mathrm{rel}}
\newcommand{\Erel}{\Etri}
\newcommand{\wErel}{\widetilde{E}_{\mathrm{3B}}}

\newcommand{\md}{\mathcal{J}}
\newcommand{\U}{\mathcal{U}}

\newcommand{\psirel}{\psi^\rel_1}
\newcommand{\LagP}[1]{L_{#1}^{(\frac12)}}
\newcommand{\Ei}{\mathrm{Ei}}

\newcommand{\sya}{\tau}
\newcommand{\syb}{t}
\newcommand{\syc}{\gamma}
\newcommand{\syd}{\nu}
\newcommand{\sye}{n}
\newcommand{\syf}{\alpha}
\newcommand{\syg}{\eta}

\begin{document}

\title{Production of three-body Efimov molecules in an optical lattice}

\author{Martin Stoll}
\affiliation{Institut f{\"u}r Theoretische Physik, Universit{\"a}t
  G{\"o}ttingen, Friedrich-Hund-Platz 1, 37077 G{\"o}ttingen, Germany}
\author{Thorsten K{\"o}hler}
\affiliation{Clarendon Laboratory, Department of Physics, University
  of Oxford, Oxford OX1 3PU, United Kingdom}
\date{\today}

\begin{abstract}
  We study the possibility of associating meta-stable Efimov trimers from 
  three free Bose atoms in a tight trap realised, for instance, via an optical 
  lattice site or a microchip. The suggested scheme for the production of 
  these molecules is based on magnetically tunable Fesh\-bach resonances and 
  takes advantage of the Efimov effect in three-body energy spectra. Our 
  predictions on the energy levels and wave functions of three pairwise 
  interacting \Rb\ atoms rely upon exact solutions of the Faddeev equations 
  and include the tightly confining potential of an isotropic harmonic atom 
  trap. The magnetic field dependence of these energy levels indicates that 
  it is the lowest energetic Efimov trimer state that can be associated in an 
  adiabatic sweep of the field strength. We show that the binding energies 
  and spatial extents of the trimer molecules produced are comparable, in 
  their magnitudes, to those of the associated diatomic Fesh\-bach molecule. 
  The three-body molecular state follows Efimov's scenario when the pairwise 
  attraction of the atoms is strengthened by tuning the magnetic field 
  strength.
\end{abstract}

\pacs{34.50.-s,36.90.+f,03.75.Lm,21.45.+v}

\maketitle

\section{Introduction}
Since the early days of quantum mechanics, the understanding of the 
complexity of few-body energy spectra has been the subject of numerous 
theoretical and experimental studies. Already in 1935 Thomas \cite{Thomas35}
predicted that three particles may be rather tightly bound even when their 
short ranged pairwise interactions support only a single, arbitrarily weakly 
bound state. The energy of the tightly bound three-body state was found to 
diverge in the hypothetical limit of a zero range binary potential. Thomas' 
discoveries were later generalised by Efimov \cite{Efimov70}, predicting that 
the number of bound states of three identical Bosons increases beyond all 
limits when the pairwise attraction between the particles is weakened in such 
a way that the only two-body bound state ceases to exist. Such an excited 
three-body energy state that appears under weakening of the attractive 
pairwise interaction is called an Efimov state. Conversely, under 
strengthening of the attractive part of the binary 
potential an Efimov state disappears into the continuum. This remarkable 
quantum phenomenon of three-body energy spectra is usually referred to as 
Efimov's scenario.

The existence of Efimov states in nature has still not been finally confirmed. 
Likely candidates may be found among the systems of identical Bosons, 
whose binary interaction potential supports only a single, weakly 
bound state. Already in 1977 Lim {\em et al.}~\cite{Lim77} predicted the 
existence of an excited state of the helium trimer molecule $^4$He$_3$ which 
followed Efimov's scenario. This discovery has later been confirmed by  
independent theoretical studies 
\cite{Cornelius86,Esry96,Kolganova98,Nielsen98,Gonzalez99,Blume00,Barletta01,Jensen04} 
using more accurate helium dimer potentials. From the experimental viewpoint, 
the observations of Ref.~\cite{Grisenti00} clearly reveal that the helium 
dimer $^4$He$_2$ is indeed weakly bound. Even the helium trimer molecule has 
been detected by diffracting a helium molecular beam \cite{Schoellkopf94} from 
a micro-fabricated material transmission grating. While state selective 
diffraction experiments with helium trimers may, in principle, be possible 
\cite{HK00}, there is still no conclusive evidence for the existence of their 
exited state.

The possibility of manipulating the low energy inter-atomic interactions, 
using magnetically tunable Fesh\-bach resonances, has provided new 
perspectives for the observation of Efimov's effect. Recent experiments with 
cold gases of Fermionic atoms \cite{Regal03,Strecker03} as well as Bosonic 
species \cite{Herbig03,Duerr04,Xu03,Thompson04} have demonstrated that 
adiabatic sweeps of the magnetic field strength can be used to associate 
highly excited diatomic Fesh\-bach molecules with an arbitrarily weak bond. 
While all these experiments were performed in atom traps under comparatively 
weak spatial confinement, there have been suggestions to produce molecules in 
the tightly confining light potential of an optical lattice \cite{Jaksch02}. 
Since tight lattices suppress number fluctuations between different sites 
\cite{Greiner02}, the molecular association may, in principle, be performed 
with two or even three atoms per site. A similarly tight or even stronger 
harmonic confinement of atoms may be achieved in microchip traps \cite{Long03}.
The energy levels of a pair of interacting atoms in a tight micro-trap have 
been determined in Refs.~\cite{BERW_FP28,Tiesinga00,Bolda02}. The universal 
properties \cite{Yamashita03,Braaten04} of three-body energy spectra in the 
presence of a confining harmonic potential have been studied in 
Ref.~\cite{Jonsell02}, using an adiabatic approximation to solve the 
stationary Sch\"odinger equation in hyper-spherical coordinates 
\cite{Blume02}. 

In this paper we exactly solve the Faddeev equations \cite{F_JETP12} to 
determine the magnetic field dependence of the energy levels of three 
identical Bose atoms, whose pairwise interactions are tuned, using the 
technique of Fesh\-bach resonances, in a harmonic micro-trap realised by an 
optical lattice site or a microchip. Our results show that linear sweeps of 
the magnetic field strength can be used to populate the lowest energetic 
meta-stable Efimov trimer molecular state, largely in analogy to the 
association of diatomic Fesh\-bach molecules. We show that the properties of 
the trimers produced, with respect to the strength of their bonds, are 
comparable to those of the associated diatomic Fesh\-bach molecule. We 
illustrate all our general findings for the example of the 155 G Fesh\-bach 
resonance of \Rb\ Bose atoms.

The paper is organised as follows: In Section \ref{Sec:two_body} we discuss 
those universal properties of near resonant diatomic bound states that are
crucial for the association of Efimov trimers. We then introduce the 
particular requirements on the type of Fesh\-bach resonance, under which 
universality can be attained over a wide range of magnetic field strengths. 
Our results indicate that broad, entrance channel dominated Fesh\-bach 
resonances may be best suited to produce the trimers and preserve their 
stability on time scales sufficiently long to study their properties.
We provide a general criterion that shows why the 155 G resonance of \Rb\ 
meets these requirements \cite{KGB03,KGG04,Marcelis04}.

In Section \ref{Sec:Thomas_and_Efimov} we first illustrate the occurrence of
the Thomas and Efimov effects in the energy spectra of three Bose atoms in 
free space, whose binary interactions are tuned using the technique of 
Fesh\-bach resonances. Our discussion reveals, in particular, why all Efimov 
trimer states in such systems are, in general, intrinsically meta-stable. We 
then show how the three-body energy spectrum is modified due to the presence 
of the harmonic confining potential of an isotropic atom trap. These 
considerations allow us to identify the lowest energetic Efimov state as 
the molecular trimer state that can be associated in an adiabatic sweep of the 
magnetic field strength.

Section \ref{Sec:trimers_in_lattices} illustrates the energy levels and wave
functions of three \Rb\ atoms under the tight confinement of an optical 
lattice site or a microchip trap. We show that, under realistic conditions, 
the trimer molecules produced, when released from the lattice, are 
sufficiently confined in space that they can be identified as separate 
entities of a dilute gas. We then suggest a general scheme for their 
detection that directly takes advantage of the periodic nature of an optical 
lattice.
 
All the details of our calculations are given in the appendices:
Appendix \ref{App:Separable_potential} introduces the separable binary 
potential \cite{L_PR135,Sandhas72,Beliaev90} that we have used to accurately 
describe the low energy scattering properties of a pair of \Rb\ atoms. We 
show, furthermore, how the separable potential approach can be extended to 
determine the two-body energy levels, in the presence of an isotropic harmonic 
atom trap, over a wide range of trap frequencies. We apply these techniques in 
Appendix \ref{app:three_body} to derive a general scheme to exactly solve
the Faddeev equations for three pairwise interacting atoms including the
confining trapping potential. Since our approach differs considerably
from the known techniques for the solution of the Faddeev equations in free
space, we provide a detailed description of their numerical implementation.
To demonstrate its predictive power with respect to three-body energy spectra, 
we provide estimates of the accuracy of our separable potential approach 
through comparisons with {\em ab initio} calculations \cite{Barletta01} of the 
helium trimer ground and excited state energies.

\section{Energy levels of a trapped atom pair}
\label{Sec:two_body}
In this section we introduce the concept of universal 
two-body scattering and bound state properties characteristic 
for cold collision physics in the vicinity of zero energy 
resonances. These universal properties are crucial for 
the existence of the Thomas and Efimov effects in three-body 
energy spectra. We describe the conditions under which the  
universality of binary physical observables can be attained 
over a wide range of magnetic field strengths in experiments 
using Fesh\-bach resonances to tune the inter-atomic 
interactions. We provide the relevant physical parameters of 
the microscopic binary potential that determine 
the energy spectra of an atom pair in free space as well as 
under the spatial confinement of an atom trap. We then show 
how the variation of the energies under adiabatic changes of 
the magnetic field strength can be used to associate diatomic 
molecules. Throughout this paper we discuss applications for 
the example of the 155 G Fesh\-bach resonance of \Rb\ 
($1\,\mathrm{G}=10^{-4}$\,T). 
The underlying physical concepts, however, are quite general 
and can be applied to cold collisions of other species of Bose 
atoms involving what we shall identify as entrance channel 
dominated resonances.

\subsection{Resonance enhanced scattering}
\subsubsection{Magnetic field tunable Fesh\-bach resonances}
Binary collisions in cold gases involve large de Broglie 
wavelengths which typically very much exceed all length scales 
set by the inter-atomic interactions. At the low collision energies 
the microscopic potential enters the description of 
scattering phenomena only in terms of a single length scale, 
the $s$ wave scattering length $a$. The experimental technique 
of Fesh\-bach resonances employs a homogeneous magnetic 
field of strength $B$ to manipulate the scattering length, taking 
advantage of the fact that the pairwise interaction 
depends on the coupling between the atomic Zeeman levels. 
Each Zeeman state is determined by the pair of total angular 
momentum quantum numbers $(f,m_f)$ of the  
hyperfine level with which the Zeeman state correlates 
adiabatically at zero magnetic field. In our applications to cold 
gases of \Rb\ the atoms are prepared in the magnetically 
trapped hyperfine state with the total angular momentum 
quantum number $f=2$ and the orientation quantum 
number $m_f=-2$ with respect to the direction of the magnetic field. 
The interaction between the atoms is usually described in 
terms of the binary scattering channels associated 
with the pairs of Zeeman levels of the individual atoms. The 
relative energies between the dissociation thresholds associated 
with the channels can be tuned using the Zeeman effect. 
We shall denote the open $s$ wave channel of a pair of 
asymptotically separated atoms of the gas as the entrance 
channel. 

The typically weak inter-channel coupling can be grossly enhanced 
by tuning the energy $E_\mathrm{res}(B)$ of a closed channel 
vibrational state $\ket{\psi_\mathrm{res}}$ (the Fesh\-bach 
resonance level) in the vicinity of the dissociation threshold of 
the entrance channel. General considerations 
\cite{Mies00,GKGTJ_JPB37} show that a virtual energy match 
between $E_\mathrm{res}(B)$ and the threshold leads to a zero 
energy resonance in the entrance channel, i.e. a 
singularity of the scattering length, described by the formula: 
\begin{equation}
  \label{eq:scattering_length_B_field}
  a=a_\mathrm{bg} \left(1-\frac{\Delta B}{B-B_0}\right).
\end{equation}
Here $a_\mathrm{bg}$ is usually referred to as the background scattering 
length, $\Delta B$ is the resonance width and $B_0$ is the position 
of the zero energy resonance.
 
\subsubsection{Universal properties of near resonant bound state 
wave functions}
The emergence of the zero energy resonance indicates the 
degeneracy of the binding energy $E_\mathrm{b}(B)$ of the highest 
excited vibrational multi-channel molecular bound state 
$\ket{\psi_\mathrm{b}}$ (the Fesh\-bach molecule) with the threshold 
for dissociation of the entrance channel at the magnetic field strength 
$B_0$. The magnetic field dependence of $E_\mathrm{b}(B)$ is due 
to the perturbation of the physically relevant multi-channel energy levels 
by the strong coupling between the entrance channel and the closed channel 
Fesh\-bach resonance state. The energy $E_\mathrm{b}(B)$ 
approaches the entrance channel dissociation threshold from the 
side of positive scattering lengths, while beyond the resonant 
field strength $B_0$, at negative scattering lengths, the bound state 
$\ket{\psi_\mathrm{b}}$ is transferred into a virtual state. At magnetic 
field strengths in the close vicinity of the zero energy resonance, 
all low energy binary collision properties are thus dominated by the 
properties of the bound state $\ket{\psidim}$, whose wave function becomes 
universal in the limit $a\to\infty$. This implies that the admixture of 
the Fesh\-bach resonance level to the molecular bound state 
$\ket{\psi_\mathrm{b}}$ vanishes in accordance with the following asymptotic 
formula \cite{GKGTJ_JPB37}: 
\begin{equation}
\label{eq:criterion_universal}
|\langle\psi_\mathrm{res}\ket{\psidim}|^2\underset{a\to\infty}{\sim}
1/\left(1+\frac{1}{2}\mu_\mathrm{res}\frac{a_\mathrm{bg}\Delta B}{a}
\frac{ma^2}{\hbar^2}\right).
\end{equation}
Here $\mu_\mathrm{res}=dE_\mathrm{res}/dB$ is the virtually constant 
magnetic moment of the resonance level $\ket{\psi_\mathrm{res}}$ and 
$m$ is the atomic mass. We note that, in general, the physically relevant
Fesh\-bach molecular state $\ket{\psidim}$ and the Fesh\-bach resonance level 
$\ket{\psi_\mathrm{res}}$ are considerably different with respect to 
their magnetic moments and their spatial extents. The state $\ket{\psidim}$
binds the atoms, while $\ket{\psi_\mathrm{res}}$ may have only a short 
lifetime, in particular, when the inter-channel coupling is strong.

In the vicinity of the magnetic field strength $B_0$ 
the long range molecular bound state $\ket{\psidim}$ consists mainly of its 
component in the entrance channel, which is given by the usual form of a 
near resonant bound state wave function \cite{KGJB03,GKGTJ_JPB37}:
\begin{equation}
   \psidim(r)\approx\frac{e^{-r/a}}{r\sqrt{2\pi a}}.
   \label{eq:near_resonant_dimer_wave_function}
\end{equation}
Its mean inter-atomic distance, i.e.~the bond length, then diverges like
the scattering length in accordance with the relationship:
\begin{equation}
  \langle r\rangle=\int\ d^3r\ r \left| \psidim(r) \right|^2\approx a/2.
  \label{eq:near_resonant_bond_length}
\end{equation}
The associated binding energy is also determined solely in terms of 
the scattering length by the universal formula:
\begin{equation}
  E_\mathrm{b}\approx-\hbar^2/(ma^2).
  \label{eq:near_resonant_binding_energy}
\end{equation}

\subsubsection{Entrance and closed channel dominated 
Fesh\-bach resonances}
The possibility of magnetically tuning the scattering length using 
Fesh\-bach resonances is unique among all physical systems. The 
universal properties of the near resonant bound state wave function 
described by Eqs.~(\ref{eq:near_resonant_dimer_wave_function}), 
(\ref{eq:near_resonant_bond_length}) and 
(\ref{eq:near_resonant_binding_energy}), however, are not directly 
related to the inter-channel coupling and apply equally well, for 
instance, also to the deuteron in nuclear physics 
\cite{BlattWeisskopf52} and to the weakly bound helium dimer 
$^4$He$_2$ van der Waals molecule \cite{Grisenti00}. Since our 
applications to the association of Efimov trimer molecules crucially 
depend on the single channel nature of the bound state 
$\ket{\psidim}$ of the Fesh\-bach molecule, we shall briefly 
outline the requirements on the Fesh\-bach resonance that assure the 
validity of the universal considerations over a significant range of 
magnetic field strengths. To this end, we consider the general 
properties of long range alkali van der Waals molecules that are determined, 
to an excellent approximation, by the scattering length in addition to 
the asymptotic form $-C_6/r^6$ of the binary potential at large 
inter-atomic distances $r$. In accordance with Ref.~\cite{GF_PRA48}, 
we shall describe the dependence of the molecular energy level on 
the van der Waals dispersion coefficient $C_6$ in terms of a mean 
scattering length $\bar{a}$, which is given by the formula:
\begin{equation}
\label{eq:mean_scattering_length}
  \bar{a}=\frac{l_{\mathrm{vdW}}}{\sqrt{2}}\ 
  \frac{\Gamma(3/4)}{\Gamma(5/4)}.
\end{equation}
Here $l_{\mathrm{vdW}}=\frac12\left(m C_6/\hbar^2\right)^{1/4}$
is usually referred to as the van der Waals length and $\Gamma$ 
denotes Euler's gamma function. The binding energy of an alkali van der 
Waals molecule is then determined by the formula \cite{GF_PRA48}:
\begin{equation}
  \label{eq:E_2BGF}
  E_\mathrm{b}=-\hbar^2/\left[m(a-\bar{a})^2\right].
\end{equation}

As discussed in detail for the examples of $^{23}$Na and  \Rb\ in 
Ref.~\cite{KGG04}, a variety of Fesh\-bach resonances can be 
classified on the basis of the properties of their associated Fesh\-bach 
molecules: Throughout this paper, we shall denote a Fesh\-bach 
resonance as entrance channel dominated when the binding energy 
$E_\mathrm{b}(B)$ is well approximated by Eq.~(\ref{eq:E_2BGF})
in some range of magnetic field strengths about $B_0$, and, at the same 
time, the contribution of the mean scattering length $\bar{a}$ of 
Eq.~(\ref{eq:mean_scattering_length}) improves the universal estimate of 
Eq.~(\ref{eq:near_resonant_binding_energy}), i.e.: 
\begin{align}
  \left|E_\mathrm{b}+\frac{\hbar^2}{m(a-\bar{a})^2}\right|
  <\left|E_\mathrm{b}+\frac{\hbar^2}{ma^2}\right|.
  \label{conditionsinglechannel1}
\end{align}
The bound state $\ket{\psi_\mathrm{b}}$ then describes a van der Waals 
molecule which implies that its properties are determined mainly by 
the entrance channel potential rather than the resonance level. In fact, 
the general considerations of Ref.~\cite{GKGTJ_JPB37} show that the admixture 
of the closed channel resonance state $\ket{\psi_\mathrm{res}}$ to the 
Fesh\-bach molecular state $\ket{\psidim}$ is negligible as soon as 
Eq.~(\ref{eq:E_2BGF}) applies to its energy 
(cf., also, Fig.~3 of Ref.~\cite{KGG04}). 
A general criterion for the applicability of Eq.~(\ref{eq:E_2BGF}) can be 
derived from the two channel approach of Ref.~\cite{GKGTJ_JPB37} which leads 
to the following inequality \cite{Petrov04FR}:
\begin{equation}
  \label{eq:condition_ecd}
  \left|\frac{\bar{a}}{a_\mathrm{bg}}
  \frac{\hbar^2/(m\bar{a}^2)}{\mu_\mathrm{res}\ \Delta B}
  \right|\ll 1.
\end{equation}

Under the conditions of Eq.~(\ref{eq:condition_ecd}),
the reason for the suppression of the closed channel contribution to 
the Fesh\-bach molecule is the large detuning of the Fesh\-bach 
resonance level from the dissociation threshold of the entrance channel 
at the position $B_0$ of the zero energy resonance \cite{GKGTJ_JPB37}. 
The parameters for the 155 G Fesh\-bach resonance of \Rb, 
i.e.~$\mu_\mathrm{res}/h=-3.12\,$MHz/G
\cite{Kokkelmans_private04}, $C_6=4703\,$a.u.~\cite{KKHV_PRL88} 
($1\,\mathrm{a.u.}=0.095734\times 10^{-24}\,$J\,nm$^6$),
$\Delta B=10.71\,$G \cite{Claussen03}, and 
$a_\mathrm{bg}=-443\,a_0$ ($a_0=0.052918\,$nm) \cite{Claussen03}, give
the quantity in Eq.~(\ref{eq:condition_ecd}) to be $4\times 10^{-2}$.
The 155 G Fesh\-bach resonance of \Rb\ is therefore entrance channel 
dominated. Similar conclusions were reached in Refs.~\cite{KGB03,Marcelis04}.  

There is also a variety of closed channel dominated Fesh\-bach 
resonances, like, e.g., those in $^{23}$Na \cite{KGG04}, whose 
Fesh\-bach molecular states $\ket{\psidim}$ become 
universal [cf.~Eq.~(\ref{eq:near_resonant_dimer_wave_function})]
only in a small region of magnetic field strengths about the zero energy
resonance at $B_0$. These states are not intermediately transferred into a 
van der Waals molecule away from the resonance. Closed channel dominated 
resonances do not satisfy the criterion of Eq.~(\ref{eq:condition_ecd}) 
and the bound states $\ket{\psidim}$ are thus significantly influenced 
by the resonance level $\ket{\psi_\mathrm{res}}$ immediately outside 
the region of universality. Our considerations on three-body energy 
spectra take into account just a single scattering channel. We shall 
therefore focus in the following just on the entrance channel 
dominated Fesh\-bach resonances of alkali Bose atoms. 

\subsection{Adiabatic association of diatomic molecules in a tight 
atom trap}
\label{subsec:adiabatic_association_2B}
Recent experiments have demonstrated the possibility of producing 
translationally cold diatomic Fesh\-bach molecules using linear downward 
ramps of the Fesh\-bach resonance level across the dissociation 
threshold of the colliding atoms. These studies on molecular 
association have been performed in quantum degenerate Fermi
gases \cite{Regal03,Strecker03}, consisting of an incoherent mixture of two 
spin states, as well as in dilute vapours of cold Bose atoms 
\cite{Herbig03,Duerr04,Xu03,Thompson04}. 

The highly excited 
Fesh\-bach molecules produced can be quite unstable with respect to 
de-excitation upon collisions with surrounding atoms. Exact calculations
of the de-excitation rate constants are challenging and have been performed 
only for transitions between tightly bound states \cite{Soldan02}. Most of 
the current knowledge, therefore, relies upon experimental evidence. For 
Fermionic species it has been predicted \cite{Petrov04} that the de-excitation 
mechanism is particularly efficient when the bond length of the Fesh\-bach 
molecule is sufficiently small for its wave function to have a significant 
spatial overlap with more tightly bound molecular states. The experimental 
studies of Ref.~\cite{Regal04} confirm this trend. 

The observation of 
large collisional de-excitation rate constants of Fesh\-bach molecules 
consisting of Bose atoms has been reported in Ref.~\cite{Mukaiyama04}. 
The $^{23}$Na resonance studied in these experiments, however, is closed 
channel dominated \cite{KGG04}. The properties of the $^{23}$Na$_2$ 
Fesh\-bach molecules of Ref.~\cite{Mukaiyama04} are thus rather different 
from those that we shall discuss in the following applications \cite{KGG04}.
The general experimental trends for both the Fermionic \cite{Regal03} and the
Bosonic \cite{Thompson04} species suggest that broad, entrance channel 
dominated Fesh\-bach resonances may be best suited to associate a large 
portion of the atoms to Fesh\-bach molecules and stabilise them. Most of the
currently known entrance channel dominated Fesh\-bach resonances have been 
found, however, in Fermionic gases \cite{Regal03,Bartenstein04}.

Inelastic de-excitation collisions with background atoms may be efficiently 
suppressed when the molecules are produced in the tight 
micro-traps of an optical lattice with an average occupation of two or three
atoms per site, respectively. Since the production of diatomic Fesh\-bach 
molecules and Efimov trimers in tight harmonic atom traps can be 
performed in similar manners, we shall first discuss the underlying physical 
concept for the simpler case of the association of a pair of atoms.  

\begin{figure}[htbp]
  \includegraphics[width=\columnwidth,clip]{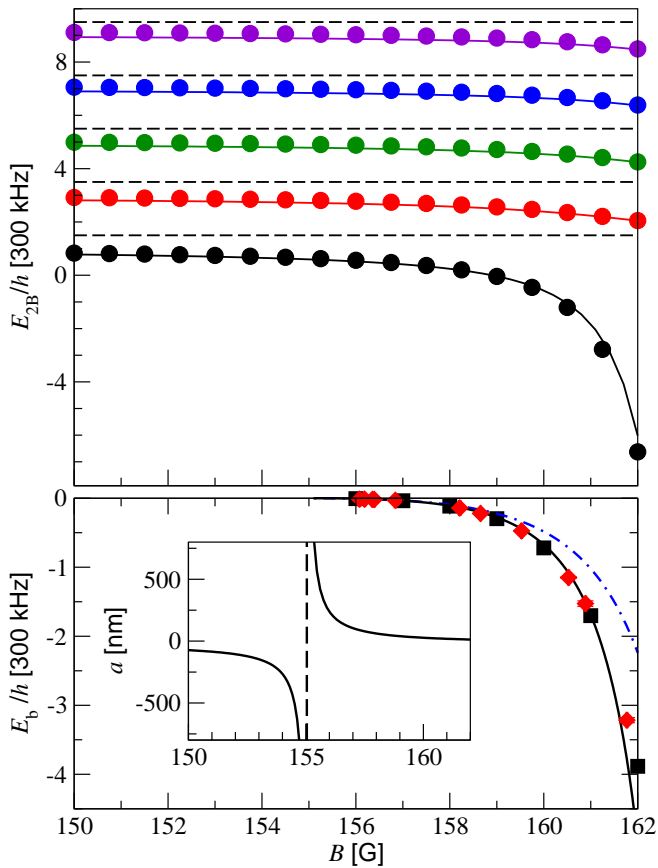}
  \caption{(Color online) Magnetic field dependence of the vibrational 
    energy levels $\Edim$ of a pair of \Rb\ atoms in 
    a $\nu_\mathrm{ho}=300$\,kHz trap (upper part) as 
    compared to the binding energies of the highest excited 
    vibrational state of the \Rbdimer\ dimer molecule in free 
    space (lower part). The circles in the upper part
    indicate numerical solutions of the two-body Schr\"odinger 
    equation for a microscopic interaction potential explicitly
    incorporating the exact scattering length as well as the 
    exact asymptotic $-C_6/r^6$ interaction energy, while
    the solid curves indicate calculations using the separable 
    potential approach of Appendix \ref{App:Separable_potential}. 
    The diamonds in the lower part are experimental binding 
    energies obtained from Ref.~\cite{Claussen03} and the 
    squares indicate results of full coupled channels calculations 
    of S. Kokkelmans \cite{Kokkelmans_private04}. The solid curve 
    indicates the dimer binding energies obtained from the 
    separable potential approach, while the dashed dotted curve 
    corresponds to their near resonant approximation 
    of Eq.~(\ref{eq:near_resonant_binding_energy}). The inset of the 
    lower part of the figure shows the singularity of the scattering 
    length at the magnetic field strength 
    $B_0=155.041$\,G \cite{Claussen03}.}
  \label{fig:dimer_energies_300kHz}
\end{figure}

The association of diatomic Fesh\-bach molecules using magnetically
tunable Fesh\-bach resonances is closely related to the variation of the 
two-body energy spectrum under adiabatic changes of the magnetic field
strength. Figure \ref{fig:dimer_energies_300kHz} shows such an 
energy spectrum versus the magnetic field strength $B$ for a pair of 
\Rb\ atoms in a tight spherically symmetric harmonic atom trap, whose high
frequency of $\nu_\mathrm{ho}=300$ kHz may be realised in a microchip trap
\cite{Westbrook} or in an optical lattice \cite{Tiesinga00}. 
The details of the underlying calculations are explained in 
Appendix \ref{App:Separable_potential}.
In Fig.~\ref{fig:dimer_energies_300kHz} the binding energy of the Fesh\-bach 
molecule in free space is shown for comparison. The spherical symmetry of 
the atom trap allows us to separate the centre of mass from the relative 
coordinates of the atoms \cite{BERW_FP28,Tiesinga00,Bolda02}, and only the 
relevant levels of the relative motion are depicted in 
Fig.~\ref{fig:dimer_energies_300kHz}. We have chosen the zero of energy, for 
each magnetic field strength, at the dissociation threshold of the entrance 
channel in free space. Following the adiabatic curves of the 
energies clearly reveals that a pair of \Rb\ atoms occupying the level closest 
to the threshold on the low field side of the zero energy resonance is 
transferred into the level of the Fesh\-bach molecule when the magnetic 
field strength is varied adiabatically across the resonance. The energy of 
the trapped molecular level and the free space binding energy $E_\mathrm{b}$ 
approach one another as the magnetic field strength is increased. A similar 
statement applies to their wave functions. This reflects the physical concept
of the adiabatic association of diatomic molecules in tight atom traps.

The two-body trap level closest to the dissociation threshold can, in 
principle, be prepared by loading a Bose-Einstein condensate adiabatically 
into an optical lattice. In the case of \Rb\ the negative background 
scattering length of $a_\mathrm{bg}=-443\ a_0$ prevents such 
a condensate from being stable on the low field side of the resonance. To 
produce \Rbdimer\ Fesh\-bach molecules via an adiabatic sweep of the 
magnetic field strength, the Bose-Einstein condensate needs to be prepared 
on the high field side of the resonance before it is loaded into the lattice. 
The resonance then needs to be crossed as quickly as possible to reach its low 
field side and, at the same time, avoid a significant heating of the atomic 
cloud \cite{KGG04}. 
Such a sequence of magnetic field sweeps is described in 
Ref.~\cite{Thompson04} in the context of the adiabatic association of 
\Rbdimer\ Fesh\-bach molecules in a cold gas.

\section{Thomas and Efimov effects}
\label{Sec:Thomas_and_Efimov}
In this section we discuss the Thomas and Efimov 
effects in the three-body energy spectra of interacting Bose atoms. 
We show that these phenomena occur when the binary scattering 
length is tuned by a magnetic field in the vicinity of a zero energy 
resonance. We then describe how the Efimov spectrum is 
modified in the presence of a trapping potential. Our results indicate that
it is the lowest energetic Efimov trimer state that can be populated by 
adiabatic changes of the magnetic field strength.

\subsection{Three-body energy levels in free space}
Throughout this section, we consider three Bose atoms that interact pairwise
through their binary potential. This assumption is justified 
for the description of weakly bound molecules, like the Efimov trimers in the
present applications, whose inter-atomic separations very much exceed the 
van der Waals length. Tightly bound trimer molecules may be significantly 
influenced by genuinely three-body forces, which we shall neglect in the 
following. We assume furthermore that the binary potential supports at most 
a single bound state, the Fesh\-bach molecule $\ket{\psidim}$. 
We thus neglect all tightly bound diatomic states, whose 
spatial extents are typically much smaller than the van der Waals length. 
In view of the large separation of the length scales between Efimov trimers
and the Fesh\-bach molecule on the one hand and the tightly bound dimer 
states on the other hand, we believe that this treatment provides an 
excellent approximation to the three-body states and their low energies we 
consider in this paper. We note, however, that any trimer state 
with an energy above the two-body ground level can, at least in principle, 
decay into a dimer bound state and a third free atom in accordance with energy 
conservation. As, in free space, the binding energies of three atoms 
are thus strictly limited, from above, by the two-body ground state energy, 
the trimer molecules under consideration are all in meta-stable states. Their 
associated lifetimes may depend sensitively on the details of the binary and 
three-body interactions. 

Given that the binary interactions support just a single, arbitrarily weakly 
bound state, it has been predicted, in terms of a rigorous variational 
treatment by Thomas \cite{Thomas35}, that three particles can be 
comparatively tightly bound. The three-body ground state can persist even 
when the binary interactions are weakened in such a way that their only 
bound state ceases to exist. Such three-body bound states that exist in the 
absence of any bound two-body subsystem are usually referred to as 
Borromean states \cite{Borromean}. 

The Thomas scenario of Borromean states has been subsequently 
generalised by Efimov \cite{Efimov70}, in a striking way, predicting that 
the number of three-body bound states of identical Bosons increases 
beyond all limits when the energy of the 
only two-body bound state is tuned towards the dissociation threshold. 
Efimov's effect is closely related to the spatial extent of the near resonant 
two-body bound state wave function, which is determined by the scattering 
length in accordance with Eq.~(\ref{eq:near_resonant_bond_length}). 
According to Efimov's treatment, it is indeed the scattering length, rather 
than the range of the potential, that sets the scale of the range of the 
effective three-particle interactions at the low collision energies under 
consideration. When the two-body binding energy reaches the dissociation 
threshold these interactions therefore acquire a long range. In contrast 
to a short range binary potential, however, long range interactions can 
support infinitely many bound states \cite{Schwinger61}. All these Efimov 
states are spherically symmetric and their energies accumulate at the
three-body dissociation threshold. Their number is predicted to follow, in the 
limit $|a|\to\infty$, the asymptotic relationship:
\begin{equation}
  N_\mathrm{Efimov}\gtrsim\frac{1}{\pi}\log\left(p_\mathrm{c} 
  \left|a\right|/\hbar\right).
  \label{eq:N_Efimov}
\end{equation}
Here $p_\mathrm{c}$ is a momentum parameter related to the range of the binary 
interactions. Efimov's remarkable results have been subsequently confirmed by 
Amado and Noble \cite{Amado72}. 

\begin{figure}[htbp]
  \includegraphics[width=\columnwidth,clip]{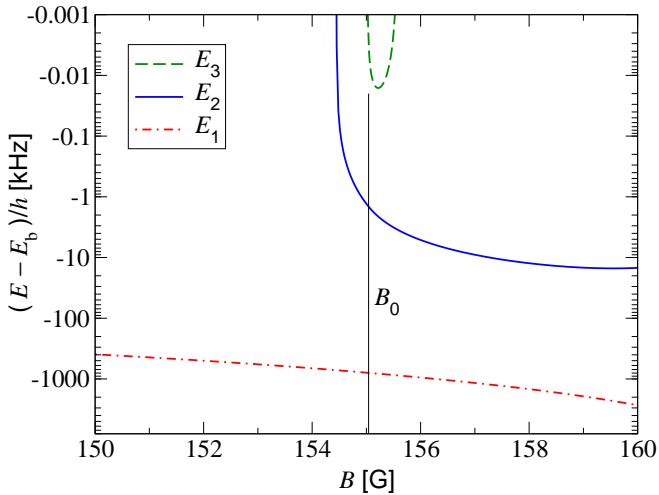}
  \caption{(Color online) Magnetic field dependence of the vibrational energy 
    levels of \Rbtrimer\ trimers relative to the binding energy $E_\mathrm{b}$ 
    of the \Rbdimer\ Fesh\-bach molecule (using $E_\mathrm{b}=0$ for $B<B_0$) 
    in free space 
    (cf.~Fig.~\ref{fig:dimer_energies_300kHz}). The first Efimov states, 
    whose energies are indicated by the solid and dashed curves, emerge at 
    about $154.4$\, G and $155$\,G, respectively. The energies of the other 
    Efimov states are not resolved even on the logarithmic energy scale. The 
    second Efimov state (dashed curve) ceases to exist at about 155.5\,G.}
  \label{fig:Efimov_free}
\end{figure}

Efimov's scenario can be realised, using the technique of Fesh\-bach 
resonances, by magnetically tuning the binary scattering length of three Bose 
atoms in the vicinity of a zero energy resonance. Figure \ref{fig:Efimov_free} 
shows the energy levels of three \Rb\ atoms in free space versus the magnetic 
field strength in the vicinity of the 155 G Fesh\-bach resonance. The exact
three-body binding energies have been determined using the momentum 
space Faddeev approach \cite{Gloeckle83} and the separable binary potential
of Appendix \ref{App:Separable_potential}.
According to Eq.~(\ref{eq:N_Efimov}), the number of Borromean Efimov 
states at negative binary scattering lengths increases beyond all limits when 
the pairwise attraction is strengthened in such a way that the two-body bound 
state emerges at the dissociation threshold. Two of their energies, i.e.~the 
solid and dashed curves, are resolved on the logarithmic scale of 
Fig.~\ref{fig:Efimov_free}. The dotted dashed curve is associated with the 
energy of the comparatively tightly bound Borromean state predicted by 
Thomas \cite{Thomas35} on the low field side of the zero energy resonance 
at $B_0=155.041$\,G (vertical solid line). 

It turns out that, as the bond of the dimer state is strengthened any further, 
the energies of the Efimov states successively cross the two-body binding 
energy and become unbound. Beyond the crossing point, the Efimov states 
can decay into a bound two-body subsystem, i.e.~the Fesh\-bach molecule, 
and a free particle, in accordance with energy conservation. This explains 
why the number of three-body bound states decreases as the attractive 
pairwise interactions are strengthened in the presence of a two-body bound 
state. In Fig.~\ref{fig:Efimov_free} we have chosen the zero of energy to 
be the three-body dissociation threshold, i.e.~the binding energy 
$E_\mathrm{b}$ of the Fesh\-bach molecule. The energy $E_3$ of the 
second Efimov state thus crosses the dimer binding energy at a magnetic field 
strength of 155.5 G. 

Following the rigorous proof of Efimov's effect by Amado and Noble 
\cite{Amado72}, the parameter $p_\mathrm{c}$ of Eq.~(\ref{eq:N_Efimov}) may 
be estimated, using the separable potential approach to the low energy 
spectrum of a pair of alkali atoms of Appendix \ref{App:Separable_potential}, 
to be:
\begin{equation}
  p_\mathrm{c}\approx 2\hbar/(\pi\bar{a}).
  \label{momentumcutoff}
\end{equation}
In agreement with Thomas' \cite{Thomas35} and Efimov's \cite{Efimov70} 
original suggestions, Eq.~(\ref{momentumcutoff}) recovers the order of 
magnitude of $\hbar/r_\mathrm{eff}$, where 
$r_\mathrm{eff}=\frac{1}{3}[\Gamma(1/4)/\Gamma(3/4)]^2\bar{a}\approx 
2.9\times \bar{a}$ is the effective range of the interaction 
between a pair of alkali atoms in the limit of large scattering lengths 
\cite{Flambaum99,Gao00}.
Equation (\ref{momentumcutoff}) thus confirms that, in contrast to the 
associated two-body problem, the low energy physics of three Bose atoms 
crucially depends on the range of the binary potential. In fact, in the 
hypothetical limit of a zero range potential not only the number of three-body 
bound states becomes infinite but also the three-body ground state energy 
diverges \cite{Thomas35}. This singular behaviour clearly reveals that the low 
energy three Boson problem is unsuited for a treatment in terms of pairwise 
contact interactions in the absence of energy cutoffs.

\subsection{Adiabatic association of Efimov trimers in an atom trap}
\begin{figure}[htbp]
  \includegraphics[width=\columnwidth,clip]{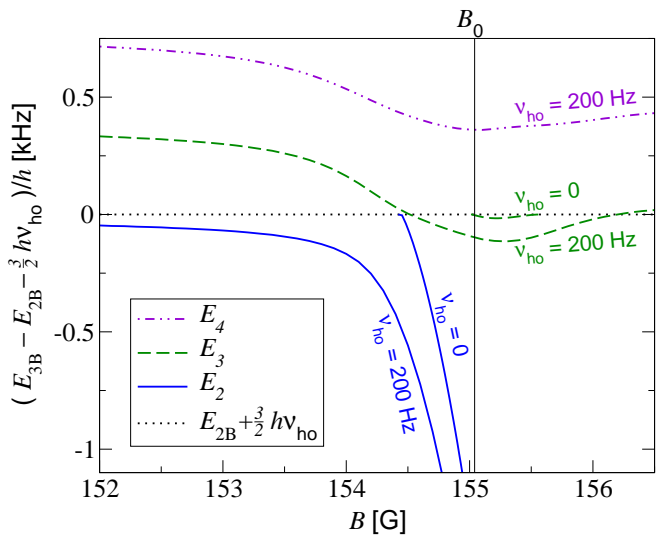}
  \caption{(Color online) Magnetic field dependence of the vibrational 
    energy levels $\Etri$ of three interacting \Rb\ atoms in a low frequency 
    $\nu_\mathrm{ho}=200$\,Hz 
    trap as compared to the energies of the free space Efimov states of 
    Fig.~\ref{fig:Efimov_free}. The energies of the trapped atoms are 
    shown relative to $\Edim+\frac{3}{2}h\nu_\mathrm{ho}$, where we have 
    chosen $\Edim$ as the lowest energy of a pair of trapped \Rb\ atoms, 
    i.e.~the two-body level that correlates adiabatically, in the limit 
    $\nu_\mathrm{ho}\to 0$, with the binding energy $E_\mathrm{b}$ of the 
    Fesh\-bach molecule of Fig.~\ref{fig:dimer_energies_300kHz}. In analogy,
    we have subtracted the dimer binding energy $E_\mathrm{b}$ from the Efimov 
    trimer energies in free space for the purpose of comparison.}
  \label{fig:trimer_energies_200Hz}
\end{figure}

The spatial confinement of an atom trap restricts the bond length of the 
Fesh\-bach molecule. This, in turn, implies that the energy levels of three
trapped atoms, unlike those in free space, do not have any accumulation point 
even when the magnetic field strength is tuned across a singularity of the 
binary scattering length. Figure \ref{fig:trimer_energies_200Hz} reveals,
however, that the energy spectrum of three \Rb\ atoms depends sensitively on 
the magnetic field strength, largely in analogy to the two-body spectrum of 
Fig.~\ref{fig:dimer_energies_300kHz}. The energy levels in 
Fig.~\ref{fig:trimer_energies_200Hz} have been obtained from exact solutions
of the Faddeev equations, in the presence of a spherically symmetric trapping 
potential, using the approach of Appendix \ref{app:three_body}. 

We have chosen the low frequency $\nu_\mathrm{ho}=200$\,Hz of a typical
magnetic atom trap in Fig.~\ref{fig:trimer_energies_200Hz}, which allows us to 
directly compare the energy levels of three \Rb\ atoms in the presence of the
trap with those of the Efimov states in Fig.~\ref{fig:Efimov_free}.
This comparison shows that the trapped three-body energy level $E_2$ 
(solid curve), closest to the three-body dissociation threshold in free space 
(dotted horizontal line), correlates adiabatically, in the limit 
$\nu_\mathrm{ho}\to 0$, with the energy of the first Efimov state.
A magnetic field pulse sequence similar to the one discussed in  
Subsection \ref{subsec:adiabatic_association_2B} and in Ref.~\cite{Thompson04}
can, in principle, be used to populate the trapped three-body energy level 
associated with the solid curve in Fig.~\ref{fig:trimer_energies_200Hz} on  
the low field side of $B_0$. 
An adiabatic upward sweep of the magnetic 
field strength across the three-body zero energy resonance of 
Fig.~\ref{fig:Efimov_free} at about 154.4\,G then transfers the trapped state 
into the first Efimov state. The physical concept underlying 
the adiabatic association of Efimov trimers in atom traps is, therefore, 
completely analogous to the considerations on the production of diatomic 
Fesh\-bach molecules in Subsection \ref{subsec:adiabatic_association_2B}.

\section{Trimer molecules in an optical lattice}
\label{Sec:trimers_in_lattices}
In this section we describe three interacting \Rb\ atoms in a tight micro-trap 
of an optical lattice site or in a microchip trap with a realistic oscillator 
frequency. Our results indicate that the spatial extent of the wave function 
of the first Efimov state can be tuned in such a way that it is smaller 
than realistic mean inter-atomic separations of dilute gases. We discuss,
furthermore, how the periodicity of an optical lattice can, in principle, be 
used to detect the trimer molecules.

\subsection{Three-body energy levels and wave functions in tightly confining 
  atom traps}
The sites of an optical lattice, in general, confine the atoms much more 
tightly than usual magnetic traps for cold gases. Their harmonic frequencies 
differ by several orders of magnitude, from tens to hundreds of kHz in the 
case of a lattice \cite{Tiesinga00} or a microchip trap 
\cite{Long03,Westbrook} as compared to about $100$\,Hz for a conventional 
magnetic trap. During the course of our studies, we have calculated the 
energy levels of three \Rb\ atoms for a variety of trap frequencies 
extending from 200\,Hz in Fig.~\ref{fig:trimer_energies_200Hz} to 1\,MHz. 
Our results indicate that, despite the pronounced differences in the spatial 
confinement, all spectra follow the same trends in their dependence on the 
magnetic field strength. The variation of the trap frequencies mainly affects
the spacings between the energy levels. In fact, all our considerations
with respect to the association of Efimov trimer molecules depend just on 
the possibility of trapping exactly three atoms rather than on the tightness
of the spatial confinement.

\begin{figure}[htbp]
  \includegraphics[width=\columnwidth,clip]{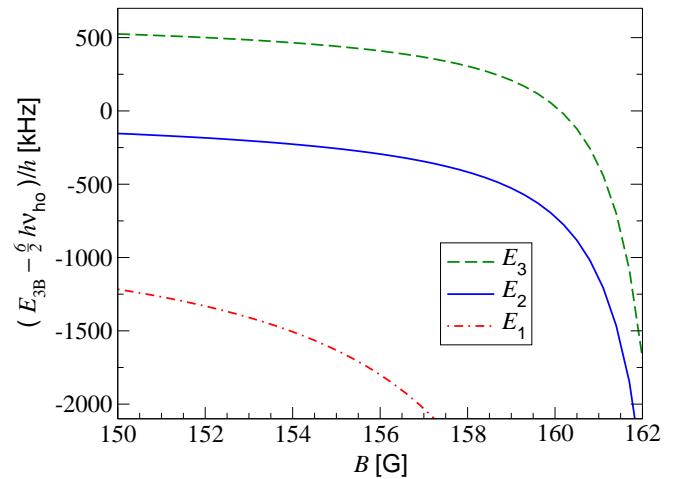}
  \caption{(Color online) Magnetic field dependence of the vibrational 
  energy levels $\Etri$ of three \Rb\  atoms relative to the zero point energy 
  $\frac62 h\nu_\mathrm{ho}$ of three 
  hypothetically non-interacting atoms in a tightly confining 
  $\nu_\mathrm{ho}=300$\,kHz trap.}
  \label{fig:trimer_energies_300kHz}
\end{figure}

As a typical example of our results, Fig.~\ref{fig:trimer_energies_300kHz} 
shows the energy levels of three \Rb\ atoms versus the magnetic field 
strength in a tightly confining atom trap with a frequency of 
$\nu_\mathrm{ho}=300$\,kHz. We note that in
Fig.~\ref{fig:trimer_energies_300kHz} the energies are given relative 
to the zero point energy of three hypothetically non-interacting trapped 
atoms, while in Figs.~\ref{fig:Efimov_free} and \ref{fig:trimer_energies_200Hz}
we have chosen the zero of energy at the three-body dissociation
threshold in free space, i.e.~at the binding energy $E_\mathrm{b}$ of the 
Fesh\-bach molecule. The solid curve of $E_2$ in 
Fig.~\ref{fig:trimer_energies_300kHz} 
thus correlates adiabatically, in the limit $\nu_\mathrm{ho}\to 0$, with the 
first Efimov state of Figs.~\ref{fig:Efimov_free} and 
\ref{fig:trimer_energies_200Hz}. 
Figure \ref{fig:Efimov_free} reveals that the trapped first Efimov state is 
transferred into a meta-stable trimer molecule, within a range of magnetic 
field strengths from 154.4\,G to at least 160\,G, when it is adiabatically 
released from the trap. The modulus of its binding energy is always slightly 
larger than $\left|E_\mathrm{b}\right|$. These predictions suggest that the 
adiabatic association of the first Efimov trimer state is, in principle, 
feasible and leads to reasonably strong bonds. We note that the production of 
the weakly bound \Rbdimer\ Fesh\-bach molecules has been observed over 
a wide range of magnetic field strengths \cite{Thompson04}.

\begin{figure}[htbp]
  \includegraphics[width=\columnwidth,clip]{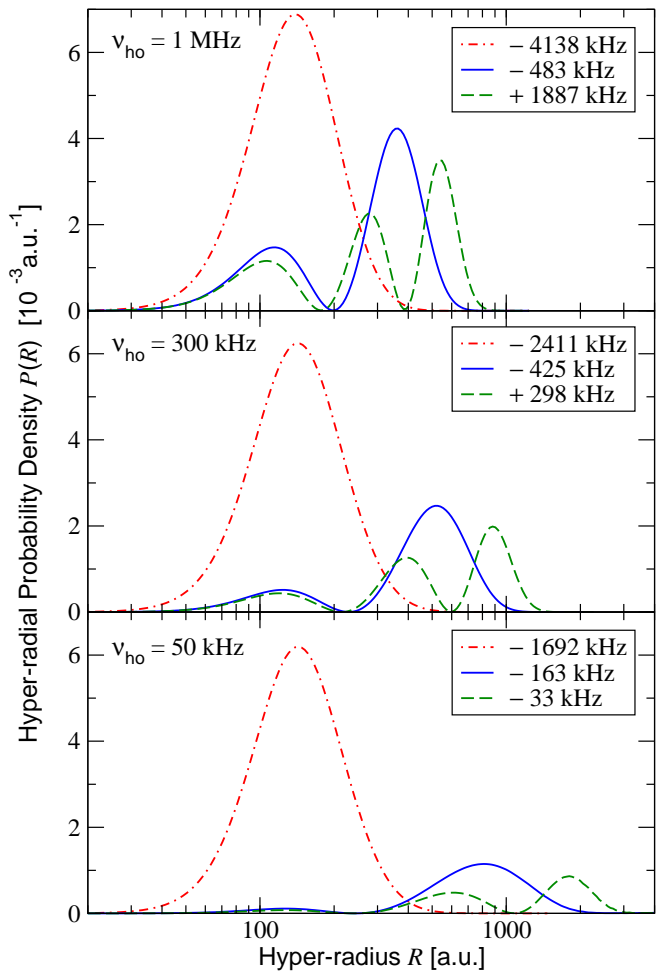}
  \caption{(Color online) Hyper-radial probability densities $P(R)$ of the 
    lowest energetic vibrational states of three \Rb\ atoms at $B=158.1$\,G 
    for the trap frequencies $\nu_\text{ho}=1\,$MHz, $\nu_\text{ho}=300\,$kHz, 
    and $\nu_\text{ho}=50\,$kHz. The legends show the associated energies 
    of the interacting atoms relative to the zero point energy of three 
    hypothetically non-interacting atoms, 
    i.e.~$\Etri-\frac62 h\nu_\mathrm{ho}$.  The hyper-radius is given on a 
    logarithmic scale.}
  \label{fig:trimer_PR_300kHz_158.1G}
\end{figure}

Weakly bound molecules, such as Efimov trimers, are characterised 
by a large spatial extent of their wave functions \cite{KGJB03}. In order
to identify them as separate entities in a dilute gas, it is crucial for the 
size of the molecules to be much smaller than the mean spacing between 
their centres of mass. As discussed in Appendix \ref{app:three_body}, 
even the isotropic wave functions of the Efimov trimers depend on 
three parameters and can therefore not be directly visualised. To give 
an impression of the spatial structure of their wave functions, 
Fig.~\ref{fig:trimer_PR_300kHz_158.1G} shows the hyper-radial 
probability densities $P(R)$ of the three lowest energetic \Rb\ trimer 
states, at a magnetic field strength of 158.1\,G, for trap frequencies 
decreasing from $1\,$MHz down to $50\,$kHz. In analogy to the case 
of diatomic molecules, the number of zeros of $P(R)$ is related to the 
degree of excitation of the three-body energy state. The solid curves 
are associated with the first Efimov state, which can be produced through 
an adiabatic upward sweep of the magnetic field strength. The wave 
function of this state extends over a few hundreds of Bohr radii in the 
$1\,$MHz trap and completely decays at about $2000\,$a.u.~in the 
case of the $50\,$kHz trap. The sizes of the trimer states produced are thus 
comparable to the single particle oscillator lengths 
($a_\mathrm{ho}=[h/(m\nu_\mathrm{ho})]^{1/2}\approx 1000\,$a.u.~for $^{85}$Rb 
atoms in a 50\,kHz trap). This indicates that the three atoms occupy a volume 
comparable to the size of the single atom oscillator ground 
state. Although these length scales very much exceed the spatial extent of 
even the most loosely bound observed diatomic ground state molecule (the 
helium dimer \cite{Grisenti00}), realistic cold gases are usually 
sufficiently dilute that 
the wave functions of these Efimov trimers do not overlap each other once they 
are adiabatically released from the lattice (cf., e.g., Ref.~\cite{Zwierlein03}
for an estimate of the remarkably large bond lengths of $^6$Li$_2$ Fesh\-bach 
molecules produced in a dilute gas).

\subsection{Detection of Efimov trimer molecules in an optical lattice}
A variety of present day detection techniques for weakly bound Fesh\-bach 
molecules in cold gases relies upon direct rf photo-dissociation 
spectroscopy \cite{Regal03}, atom loss and recovery measurements 
\cite{Strecker03,Thompson04}, or the spatial separation of the molecular 
cloud from the remnant atomic gas \cite{Herbig03,Duerr04,Xu03}. While all
these techniques may be applicable, in one way or another, also to Efimov 
trimers, an optical lattice lends itself to a rather different approach to 
their detection; the mass spectrometry using the periodic light potential as 
a diffracting device. We shall focus in the following on the perspectives 
of the diffraction technique.

The crucial coherence properties of Bose-Einstein condensed atomic gases 
loaded into optical lattices have been studied in detail in
Ref.~\cite{Greiner02}. These experiments indicate that the possibility of 
diffraction depends on the weakness of the light potential. Contrary to this, 
the association of Efimov trimers requires high tunnelling barriers between 
the individual sites to protect the molecules against inelastic collisions. In 
accordance with Ref.~\cite{Greiner02}, we expect that adiabatically releasing 
the lattice depth transfers the gas of the Efimov trimers produced from their 
insulating phase back into the super-fluid phase. Super-fluid gases in optical 
lattices, however, can be diffracted \cite{Greiner02}. 

The spatial periodicity of the optical lattice implies that the momenta of 
the cold molecules, when released from the light potential, are determined 
by multiples of the reciprocal lattice vectors with a negligible spread
under the conditions of super-fluidity. The associated quantised velocity 
transfers obtained from the lattice, and consequently also the diffraction 
angles, are therefore inversely proportional to the molecular mass. 
The principle of this mass selection technique has been demonstrated in 
several earlier experiments on the spatial separation of weakly bound helium 
dimers \cite{Grisenti00,Schoellkopf94} and trimers \cite{Schoellkopf94} 
as well as sodium dimers \cite{Chapman95} from molecular beams. The freely 
expanding \Rb\ Efimov trimers of the present applications may then be 
dissociated and imaged using the methods of 
Refs.~\cite{Strecker03,Herbig03,Duerr04,Xu03,Thompson04}. 
The mass selection detection technique suggested in this paper is general and 
could be applied, for instance, also to diatomic Fesh\-bach molecules. 

We expect that the most serious constraint on the production and detection of 
Efimov trimers in optical lattices is their intrinsic meta-stability. In the 
special case of \Rbtrimer\ molecules there are two mechanisms that can lead to 
their spontaneous dissociation: The first mechanism consists in the decay into 
a fast, tightly bound dimer molecule and a fast atom. The second decay 
scenario involves spin relaxation of a constituent of the trimer molecule in 
analogy to the studies of Refs.~\cite{Thompson04,KTJ04}.
While the present considerations do not allow us to estimate the molecular 
lifetimes associated with the first decay mechanism, it has been shown in 
Refs.~\cite{Thompson04,KTJ04} that spin relaxation can be efficiently 
suppressed by increasing the spatial extent of the molecules. It can be 
ruled out completely for other atomic species, in which the individual atoms 
are prepared in their electronic ground state. We are, however, not aware of  
other well studied entrance channel dominated Fesh\-bach resonances of 
identical Bose atoms besides \Rb.

\section{Conclusions}
We have studied in this paper the association of weakly bound meta-stable 
trimer molecules from three free Bose atoms in the ground state of a tight 
micro-trap of an optical lattice site or of a microchip. Our approach takes 
advantage of a remarkable quantum phenomenon in three-body energy spectra, 
known as Efimov's effect. Efimov's effect occurs when the binary scattering 
length is tuned in the vicinity of a zero energy resonance and involves the 
emergence of infinitely many three-body molecular states. We have shown that 
this scenario can be realised by magnetically tuning the inter-atomic 
interactions using the technique of Fesh\-bach resonances. 

The association scheme for trimer molecules, suggested in this paper, involves 
an adiabatic sweep of the magnetic field strength across a three-body zero 
energy resonance and can be performed largely in analogy to the well known 
association of diatomic molecules via magnetically tunable Fesh\-bach 
resonances. Our results indicate that the predicted binding energies and 
spatial extents of the Efimov trimer molecules produced are comparable in 
their magnitudes to the associated quantities of the diatomic Fesh\-bach 
molecules. We have illustrated our general considerations for the example 
of \Rb\  including a suggestion for a complete experimental scenario and a 
possible detection scheme. 

Once the meta-stable trimer molecules are produced, the possibility 
of tuning the inter-atomic interactions using Fesh\-bach resonances may, 
in principle, be exploited to study the Efimov property of the trimer state. 
Since, according to the predictions of this paper, it is the first Efimov 
state that gets associated in an adiabatic sweep of the magnetic field 
strength, the trimer molecules are expected to dissociate as the pairwise 
attraction between the atoms is strengthened. This, at first sight, 
counterintuitive scenario can be realised by tuning the magnetic field 
strength away from the zero energy resonance on the side of positive 
scattering lengths. Once the energy of the diatomic Fesh\-bach molecule 
crosses the binding energy of the trimers, the Efimov states dissociate 
into the Fesh\-bach molecule as well as a third free atom. In this way, the 
technique of Fesh\-bach resonances could provide a unique opportunity to 
finally confirm this predicted \cite{Efimov70} but as yet unobserved 
fascinating quantum phenomenon in the energy spectrum of three Bose particles.

\acknowledgments
This research has been supported by the Deutsche For\-schungs\-gemein\-schaft
and by the Royal Society.

\appendix

\section{Separable two-body interaction}
\label{App:Separable_potential}
In this appendix we provide a convenient effective binary interaction 
potential that well describes the relevant low energy vibrational levels of an 
atom pair, both in free space and under the strong spatial confinement of a 
tight micro-trap of an optical lattice site or of a microchip. We 
determine the parameters of the effective potential in terms of the $s$ wave 
binary scattering length $a$ and the van der Waals dispersion coefficient 
$C_6$, which characterises the interaction energy at asymptotically large 
inter-atomic distances.

\subsection{Overview of the separable potential approach in free space}
\label{SubSecApp:Scattering_in_free_space}
\subsubsection{Hamiltonian}
We first consider a pair of identical Bose atoms of mass $m$ at the
positions $\vec r_1$ and $\vec r_2$, respectively, which interact via
the microscopic potential $V$ in the absence of a confining atom trap. The
associated free-space binary Hamiltonian is then given by the formula:
\begin{equation}
  \label{eq:hamilton_free_two_particles}
  H_\mathrm{free}=H_0^\mathrm{free} + V.
\end{equation}
Here the non-interacting Hamiltonian
\begin{equation}
  \label{eq:hamilton0_free_two_particles}
  H_0^\mathrm{free}=-\frac{\hbar^2}{2(2m)}\nabla_{\vec R}^2
  -\frac{\hbar^2}{2(\frac m2)}\nabla_{\vec r}^2
\end{equation}
accounts for the kinetic energy, where $\vec R=\frac12(\vec r_1+\vec
r_2)$ and $\vec r=\vec r_1-\vec r_2$ denote the centre of mass and
relative coordinates of the atom pair, respectively. We assume in the
following that the centre of mass is at rest and focus only on the
relative motion of the atom pair. The non-interacting Hamiltonian in free 
space then reduces to $\hat{H}_0^\mathrm{free}=-\hbar^2\nabla_{\vec r}^2/m$.

\subsubsection{Transition matrix}
In our subsequent applications to three-body systems it will be convenient to 
represent all bound and free energy levels of the binary subsystems in terms 
of their transition matrix $\hat{T}(z)$ associated with the relative motion 
of the atoms, whose singularities as a function of the continuous variable 
$z$ determine the two-body energy spectrum. In general, the transition matrix 
associated with the interaction potential $V$ can be obtained from the 
Lippmann-Schwinger equation \cite{newton}:
\begin{equation}
  \label{eq:T-Matrix}
  \hat{T}(z)=V+V\hat{G}_0^\mathrm{free}(z)\hat{T}(z).
\end{equation}
Here $\hat{G}_0^\mathrm{free}(z)=(z-\hat{H}_0^\mathrm{free})^{-1}$ is the 
Green's function of the relative motion of the atoms in the absence of 
inter-atomic interactions. 

\subsubsection{Low energy physical observables}
The solution of the two-body Lippmann-Schwinger equation (\ref{eq:T-Matrix}) 
for the microscopic inter-atomic potential $V(\vec{r})$ is a demanding problem 
in its own right. The full binary interaction, however, describes a range of 
energies much larger than those accessible to cold collision physics. We 
shall therefore introduce a simpler effective potential which properly 
accounts just for the relevant low energy physical observables. Considerations 
\cite{GKGTJ_JPB37} beyond the scope of this paper show that all physical 
observables associated with cold binary collisions can be described by a 
single parameter, the $s$ wave scattering length $a$. The $T$ matrix 
determines the scattering length by its plane wave matrix elements in the 
limit of zero energy:
\begin{equation}
  \label{eq:T-Matrix-separable_normalisation}
  \bra{\vec p'=0}\hat{T}(0)\ket{\vec p=0}=
  \frac1{(2\pi\hbar)^3}\frac{4\pi\hbar^2}{m}a.
\end{equation}

The very existence of the Thomas and Efimov effects clearly reveals that cold 
collisions of three Bose atoms are sensitive also to the spatial range of the 
interaction. At large distances $r$ the asymptotic form of the inter-atomic  
potential is given by $V(\vec{r})\underset{r\to\infty}{\sim} -C_6/r^6$, where 
$C_6$ is the van der Waals dispersion coefficient. We shall thus determine the 
effective binary potential of alkali atoms in such a way that it provides a 
straightforward access to the full $T$ matrix and, at the same time, recovers 
the exact scattering length $a$ as well as those low energy physical 
observables that are sensitive also to $C_6$, such as, for instance, the 
the energies of the trapped atom pairs in 
Fig.~\ref{fig:dimer_energies_300kHz}. As shown in Ref.~\cite{GF_PRA48}, 
the $C_6$ coefficient enters these measurable quantities in terms of the mean 
scattering length of Eq.~(\ref{eq:mean_scattering_length}).
In our applications to energy spectra in the vicinity of
a zero energy resonance the mean scattering length $\bar{a}$ determines the 
binding energy of the highest excited diatomic vibrational state 
(cf.~Fig.~\ref{fig:dimer_energies_300kHz}) at positive scattering lengths $a$ 
by Eq.~(\ref{eq:E_2BGF}).

\subsubsection{Solution of the Lippmann-Schwinger equation}
To efficiently solve the two-body and three-body Schr\"odinger
equations, it is convenient to choose a (non-local) separable potential 
of the general form (see, e.g., Ref.~\cite{L_PR135,Sandhas72,Beliaev90})
\begin{equation}
  \label{eq:separable_potential}
  V_\mathrm{sep}=\ket{g}A\bra{g}
\end{equation} 
as an effective replacement of the full microscopic binary interaction 
$V(\vec{r})$. Here $\ket{g}$ is usually referred to as the form factor 
that sets the scale of the spatial range of the potential, while the amplitude 
$A$ determines the interaction strength. For convenience, we choose the form 
factor to be a Gaussian function in momentum space
\cite{GKGTJ_JPB37}:
\begin{equation}
  \label{eq:formfactor_mom}
  g(p)=\braket{\vec p}{g}=
  \left(\frac{\sigma^2}{\pi\hbar^2}\right)^{3/4} 
  \exp\left(-\frac{p^2\sigma^2}{2\hbar^2}\right).
\end{equation}

To adjust the amplitude $A$ and the range parameter $\sigma$ in such a way 
that $V_\mathrm{sep}$ recovers the scattering length $a$ as well as the 
mean scattering length $\bar{a}$ of the microscopic interaction potential 
$V(\vec{r})$, we determine the full two-body energy spectrum associated with 
the separable potential via its transition matrix. We thus solve the 
Lippmann-Schwinger equation (\ref{eq:T-Matrix}) formally, by iteration, in 
terms of its Born series: 
\begin{equation}
  \label{eq:Born_series}
  \hat{T}_\mathrm{sep}(z)=\sum_{j=0}^\infty
  \left[
    V_\mathrm{sep}
    \hat{G}_0^\mathrm{free}(z)
    \right]^j 
  V_\mathrm{sep}.
\end{equation}
A simple derivation then shows that the $T$ matrix 
$\hat{T}_\mathrm{sep}(z)$ associated with $V_\mathrm{sep}$ is given by
the formula:
\begin{equation}
  \label{eq:T-Matrix-separable}
  \hat{T}_\mathrm{sep}(z)=\ket{g}\tau_{\mathrm{free}}(z)\bra{g}.
\end{equation}
Here the function $\tau_{\mathrm{free}}(z)$ can be determined from a geometric 
series to be:
\begin{equation}
  \label{eq:T-Matrix-denominator_tau}
  \tau_{\mathrm{free}}(z)=\left[A^{-1}
  -\bra{g}\hat{G}^{\mathrm{free}}_0(z)\ket{g}\right]^{-1}.
\end{equation}

\subsubsection{Adjustment of the separable potential}
Evaluated at zero energy, $\tau_{\mathrm{free}}(z=0)$ is related to the 
$s$ wave binary scattering length $a$ through 
Eq.~(\ref{eq:T-Matrix-separable_normalisation}). 
Given the Gaussian form of $|g\rangle$ in Eq.~(\ref{eq:formfactor_mom}), 
a spectral decomposition of the Green's function $\hat{G}_0^\mathrm{free}(z)$ 
in terms of plane wave momentum states shows that its matrix element in 
Eq.~(\ref{eq:T-Matrix-denominator_tau}) can be evaluated at zero energy to be 
$\bra{g}\hat{G}^{\mathrm{free}}_0(0)\ket{g}=-2m\sigma^2/\hbar^2$. 
This leads to the relation
\begin{equation}
  \label{eq:separablepotential_A}
  A=\frac{-\hbar^2/(2m\sigma^2)}{1-\sqrt{\pi}\sigma/a},
\end{equation}
which can be used to eliminate the amplitude $A$ in favour of the range
parameter $\sigma$ and the scattering length $a$. 

The single pole of $\tau_{\mathrm{free}}(z)$ at positive scattering lengths
determines the energy $E_\mathrm{b}$ of the highest excited near resonant 
vibrational bound state (cf.~Fig.~\ref{fig:dimer_energies_300kHz}). 
The adjustment of the energy $E_\mathrm{b}$ to Eq.~(\ref{eq:E_2BGF}) has been 
performed in Ref.~\cite{GKGTJ_JPB37} and determines the remaining unknown 
range parameter to be:
\begin{equation}
  \label{eq:range_parameter_sigma}
  \sigma=\sqrt{\pi}\bar a/2.
\end{equation}
In our application to {\Rb} we use $\sigma=69.58$\,{a.u.} which corresponds 
to $C_6=4703$\,{a.u.}
\cite{KKHV_PRL88}. The amplitude $A$ depends on the magnetic field 
strength $B$ via the scattering length $a$ in accordance with
Eq.~(\ref{eq:separablepotential_A}).

\subsection{Energy levels of a trapped atom pair}
\label{SubSecApp:Two_body_energy_levels}
\subsubsection{Energy levels of the relative motion in the absence of
  an inter-atomic interaction}
In the following applications we describe the micro-trap by a three 
dimensional spherically symmetric harmonic 
potential. The linear confining force then allows us to separate the 
centre of mass motion from the relative motion of a trapped atom pair
\cite{BERW_FP28,Tiesinga00,Bolda02}.
In the absence of an inter-atomic interaction the Hamiltonian of the 
relative motion is thus given by:
\begin{equation}
  \label{eq:hamilton_HO_one_particle}
  \hat{H}_0=-\frac{\hbar^2}{2\left(\frac{m}{2}\right)}\nabla_{\vec r}^2+ 
  \frac12 \left(\frac{m}{2}\right) \omega_\mathrm{ho}^2 \vec r^2.
\end{equation}
Here $\omega_\mathrm{ho}$ is the angular trap frequency.
Throughout this appendix we choose energy states $\ket{\varphi_{klm_l}}$ 
of the harmonic oscillator with a definite orbital angular momentum, 
where $l$ is the angular momentum quantum number and $m_l$ is the 
orientation quantum number. The associated energies are given 
by $E_{kl}=\hbar\omega_\mathrm{ho}(2k+l+3/2)$, where $k=0,1,2,\ldots$ 
labels the vibrational excitation of the atom pair \cite{morsefeshbach}. 
We denote the spherically symmetric vibrational energy states by
$\ket{\varphi_k}=\ket{\varphi_{k00}}$ and their energies by $E_k=E_{k0}$. 
Their wave functions are given by 
\begin{equation}
  \label{eq:wavefunction_HO_one_particle_sphsym_posspace}
  \varphi_k(r)=\sqrt{\frac{\beta^{3/2}}{2\pi} 
    \frac{\Gamma(k+1)}{\Gamma(k+\frac32)}}
  \ e^{-\beta r^2/2}
  \ \LagP{k}(\beta r^2),
\end{equation}
where $\LagP{k}$ is an associated Laguerre polynomial. The parameter
$\beta=m\omega_\mathrm{ho}/(2\hbar)$ is related to the harmonic
oscillator length $a_\mathrm{ho}$ for a single atom, i.e.~the trap
length, by $\beta=1/(2a_\mathrm{ho}^2)$.

\subsubsection{Separable potential approach in the presence of a trapping 
  potential}
The spherical symmetry of the trap allows us to determine the energy levels 
of the relative motion of a pair of interacting trapped atoms in complete 
analogy to their counterparts in free space 
\cite{BERW_FP28,Tiesinga00,Bolda02}. The associated Hamiltonian is then given 
by 
\begin{equation}
  \label{eq:hamilton_HO_two_particles}
  \hat{H}=\hat{H}_0+V(\vec{r}),
\end{equation}
where $V(\vec{r})$ is the spherically symmetric microscopic inter-atomic 
potential and $\hat{H}_0$ is the Hamiltonian of 
Eq.~(\ref{eq:hamilton_HO_one_particle}). In the following, we shall denote the 
energies associated with $\hat{H}$ by $\Edim$.

Given that typically the trap length 
$a_\mathrm{ho}$ very much exceeds the van der Waals length $l_\mathrm{vdW}$, 
the full microscopic interaction can be replaced by the separable potential 
of Appendix \ref{SubSecApp:Scattering_in_free_space} to describe the limited 
range of energies involved in the adiabatic association of molecules. The 
associated $T$ matrix $\hat{T}_\mathrm{sep}(z)$ of the relative motion of a 
pair of trapped interacting atoms can then be determined in analogy to 
Appendix \ref{SubSecApp:Scattering_in_free_space}. This yields:   
\begin{equation}
  \label{eq:T-Matrix-separable-1}
  \hat{T}_\mathrm{sep}(z)=\ket{g}\tau(z)\bra{g}.
\end{equation}
The function $\tau(z)$ can be obtained from 
Eq.~(\ref{eq:T-Matrix-denominator_tau}) by replacing the Green's function 
$\hat{G}_0^\mathrm{free}(z)$ in free space by its counterpart 
$\hat{G}_0(z)=(z-\hat{H}_0)^{-1}$ in the presence of the trapping potential. 
This leads to the relation:
\begin{equation}
  \label{eq:T-Matrix-denominator_tau-trap}
  \tau(z)=\left[
    \left(\tau_{\mathrm{free}}(0)\right)^{-1} 
    -4\left(\frac m2\right) \sigma^2/\hbar^2
    -\bra{g}\hat{G}_0(z)\ket{g}
  \right]^{-1}.
\end{equation}
The poles of $\tau(z)$ determine the energy levels of the Hamiltonian 
(\ref{eq:hamilton_HO_two_particles}) in the separable potential approximation,
i.e.~the poles are located at the energies $z=\Edim$. 

We have evaluated the function $\tau(z)$ in terms of the oscillator
states $\ket{\varphi_k}$ of
Eq.~(\ref{eq:wavefunction_HO_one_particle_sphsym_posspace}) using the
spectral decomposition of the Green's function $\hat{G}_0(z)$.  The
matrix element relevant to Eq.~(\ref{eq:T-Matrix-denominator_tau-trap}) is 
given by:
\begin{equation}
  \bra{g}\hat{G}_0(z)\ket{g}=\sum_{k=0}^\infty
  \frac{\left|\braket{\varphi_k}{g}\right|^2}{z-E_{k}}.
\end{equation}  
During the course of our studies, we have compared the separable
potential approach to the two-body energy spectrum to predictions
obtained with a microscopic potential $V(\vec{r})$ for a variety
of scattering lengths and trap frequencies. Figure
\ref{fig:dimer_energies_300kHz} shows such a comparison for a rather
tight $\nu_\mathrm{ho}=300$ kHz atom trap, which clearly reveals the
applicability of the separable potential approach in the range of
energies relevant to the adiabatic association of Efimov trimer
molecules.

\section{Three-body energy levels and wave functions}
\label{app:three_body}
In this appendix we derive the Faddeev equations that determine
the exact energy levels of three interacting Bose atoms in the confining 
potential of a spherical trap. We then describe a practical method to  
exactly solve these equations in the separable potential approach.

\subsection{Faddeev approach}
\subsubsection{Three-body Hamiltonian and Jacobi coordinates}
\begin{figure}[htbp]
  \includegraphics[width=\columnwidth,clip]{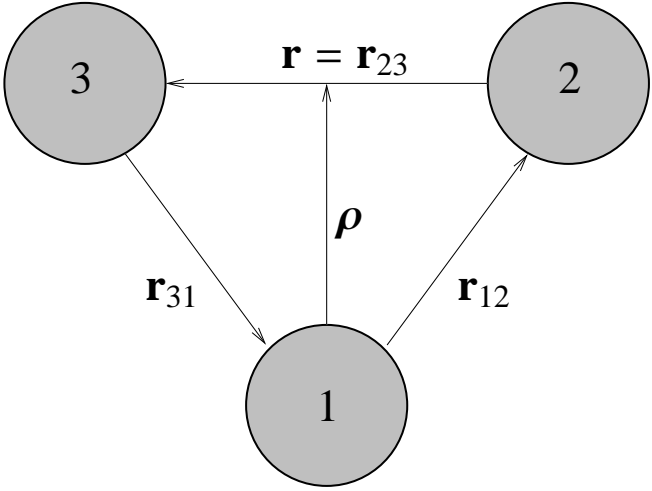}
  \caption{Jacobi coordinates of the relative motion of three atoms. 
  The set of coordinates $\vec{\rho}$ and $\vec{r}$ is selected in such a way
  that it is suited to describe the hypothetical situation of an interacting 
  pair of atoms $(2,3)$ with atom 1 playing the role of a spectator.}
  \label{fig:Jacobi}
\end{figure}

Throughout this appendix, we assume that the atoms interact pairwise
via the potential $V(\vec{r})$. The complete Hamiltonian is then given by: 
\begin{eqnarray}
  \label{eq:hamilton_HO_three_particles}
  H=H_0 +V(\vec r_{23})+V(\vec r_{31})+V(\vec r_{12}).
\end{eqnarray}
Here the non-interacting Hamiltonian
\begin{eqnarray}
  \label{eq:hamilton0_HO_three_particles}
  H_0=\sum_{i=1}^3 
  \left(-\frac{\hbar^2}{2 m}\nabla_{\vec r_i}^2+
  \frac12 m\omega_\mathrm{ho}^2\vec r_i^2\right)
\end{eqnarray}
accounts for the kinetic energy and the harmonic trapping potential of each 
atom, while $V(\vec r_{ij})$ describes the interaction between the atoms $i$ 
and $j$ ($i,j=1,2,3$) in dependence on their relative coordinates 
$\vec{r}_{ij}=\vec r_i-\vec r_j$. In the following, we employ the Jacobi 
coordinates 
$\vec R=(\vec r_1+\vec r_2+\vec r_3)/3$,
$\vec \rho=\vec r_1-(\vec r_2+\vec r_3)/2$ and
$\vec r=\vec r_2-\vec r_3$
of Fig.~\ref{fig:Jacobi} to separate out the three-body centre of mass. In 
analogy to the case of a pair of trapped atoms, the harmonic force then allows 
us to divide the non-interacting three-body Hamiltonian 
of Eq.~(\ref{eq:hamilton0_HO_three_particles}) 
into three harmonic oscillator contributions associated with the Jacobi 
coordinates. The binary potentials involve only the relative coordinates 
$\vec\rho$ and $\vec{r}$. The complete three-body Hamiltonian can thus be 
represented by:
\begin{align}
  \nonumber
  H=&-\frac{\hbar^2}{2(3m)}\nabla_{\vec{R}}^2
    +\frac12 (3m)\omega_\mathrm{ho}^2 \vec{R}^2
    -\frac{\hbar^2}{2(\frac23 m)}\nabla_{\vec{\rho}}^2\\
  \nonumber
  &+\frac12\left(\frac23 m\right)\omega_{\mathrm{ho}}^2 
  \vec{\rho}^2
  -\frac{\hbar^2}{2(\frac{m}{2})}\nabla_{\vec{r}}^2
  +\frac12\left(\frac{m}{2}\right)\omega_\mathrm{ho}^2\vec{r}^2\\
  \label{eq:hamilton_HO_three_particles_jac}
  &
  + V\!\left(\vec{r}\right)
  + V\!\left(\vec{\rho}+\frac12\vec{r}\right)
  + V\!\left(\vec{\rho}-\frac12\vec{r}\right).
\end{align}

\subsubsection{Faddeev equations for three trapped atoms}
The Faddeev equations for the energy levels of three pairwise interacting
atoms in the confining potential of a trap can be derived largely in 
analogy to their counterparts in free space \cite{F_JETP12}. To this end, 
we introduce the Green's function
\begin{equation}
  G_0(z)=(z-H_0)^{-1}
\end{equation}
associated with the non-interacting Hamiltonian of 
Eq.~(\ref{eq:hamilton0_HO_three_particles}) and denote the binary 
potential associated with the atom pair $(i,j)$ by $V_k=V(\vec r_{ij})$ for 
each of the three possible combinations of atomic indices 
\begin{displaymath}
  (ijk)=(1,2,3),(2,3,1),(3,1,2).
\end{displaymath}
The stationary Schr{\"o}dinger equation $H\ket{\psi}=E\ket{\psi}$ for a 
three-body energy state $\ket{\psi}$ can then be represented in terms of 
the matrix equation:
\begin{equation}
  \label{eq:schroedinger_equation_three_particles_trap}
  \ket{\psi}=G_0(E)\left(V_1+V_2+V_3\right)\ket{\psi}.
\end{equation}
Introducing the Faddeev components
\begin{equation}
  \label{eq:faddeev_component}
  \ket{\psi_i}=G_0(E)V_i\ket{\psi},
\end{equation}
the three-body energy state is given by
$\ket{\psi}=\ket{\psi_1}+\ket{\psi_2}+\ket{\psi_3}$. Inserting this Faddeev
decomposition into Eq.~(\ref{eq:schroedinger_equation_three_particles_trap}) 
on the left hand side and rearranging the terms yields:
\begin{equation}
  \label{eq:schroedinger_equation_three_particles_trap_psi1}
  \left[1-G_0(E)V_1\right]\ket{\psi_1}=G_0(E)V_1
    \left(\ket{\psi_2}+\ket{\psi_3}\right).
\end{equation}
Equation (\ref{eq:schroedinger_equation_three_particles_trap_psi1}) can then 
be solved formally for $\ket{\psi_1}$ by multiplying both sides with 
$\left[1-G_0(E)V_1\right]^{-1}$ from the left. This leads to the Faddeev 
equation:
\begin{equation}
  \label{eq:Faddeev1}
  \ket{\psi_1}=\left[1-G_0(E)V_1\right]^{-1}
  G_0(E)V_1\left(\ket{\psi_2}+\ket{\psi_3}\right).
\end{equation}
The kernel of Eq.~(\ref{eq:Faddeev1}) can then be expanded into the power 
series associated with the inverse matrix $\left[1-G_0(E)V_1\right]^{-1}$. 
This expansion yields:
\begin{equation}
  \label{eq:Faddeev_kernel}
  \left[1-G_0(E)V_1\right]^{-1}G_0(E)V_1=
  G_0(E)\sum_{j=0}^\infty 
  \left[
    V_1G_0(E)
    \right]^j 
  V_1.
\end{equation}
In analogy to Eq.~(\ref{eq:Born_series}), the sum on the right hand side 
of Eq.~(\ref{eq:Faddeev_kernel}) can be identified as the Born series
associated with the Lippmann-Schwinger equation:                               
\begin{equation}
  \label{eq:faddeev_equation_two-body_t-matrices}
  T_1(z)=V_1+V_1G_0(z)T_1(z).
\end{equation}
Representing Eq.~(\ref{eq:Faddeev1}) in terms of the $T$ matrix $T_1(E)$ for 
the interacting pair of atoms $(2,3)$ then recovers the Faddeev equation for
$\ket{\psi_1}$ in its original form \cite{F_JETP12}:
\begin{equation}
  \label{eq:faddeev_equation_psi1}
  \ket{\psi_1}=G_0(E)T_1(E)\left(\ket{\psi_2}+\ket{\psi_3}\right).
\end{equation}
The Faddeev equations for $\ket{\psi_2}$ and $\ket{\psi_3}$ are obtained 
by cyclic permutations of the atomic indices. The resulting set of three 
coupled matrix equations for $\ket{\psi_1}$, $\ket{\psi_2}$ and 
$\ket{\psi_3}$ determines all energy levels of three interacting atoms in 
a trap. We note that the formal derivations leading to 
Eq.~(\ref{eq:faddeev_equation_psi1}) do not refer to the specific nature of
the confining potential. This is the reason for the general validity of the 
Faddeev approach in free space as well as in the presence of an atom trap.

\subsubsection{Faddeev approach for three identical Bose atoms}
In the special case of systems of three identical Bose atoms, 
such as {\Rbtrimer}, the three Faddeev components depend on each other 
through cyclic permutations of the atoms. These permutations can be  
represented by a unitary operator $\U$, 
i.e.~$\ket{\psi_2}=\U\ket{\psi_1}$ and $\ket{\psi_3}=\U^2\ket{\psi_1}$, 
which transforms the three-body wave functions in accordance with the 
formula:
\begin{equation}
  (\U\psi)(\vec R,\vec\rho,\vec r)=\psi(\vec R,\vec\rho',\vec r').
\end{equation}
Here the primed coordinates are determined by
$\textstyle{{\vec{\rho'}}\choose{\vec{r'}}}=
\md\textstyle{{\vec\rho}\choose{\vec r}}$, where
\begin{equation}
  \label{eq:md-definition}
    \md=
    \left( \begin{array}{rr}
        -\frac12 \Eins & \frac34 \Eins\\
        -\Eins & -\frac12 \Eins
      \end{array} \right)
\end{equation}
is a $6\times6$ matrix satisfying $\md^3=1$. The first Faddeev
component of Eq.~(\ref{eq:faddeev_equation_psi1}) is thus determined 
by the single Faddeev equation:
\begin{equation}
  \label{eq:faddeev_equation_psi1-1}
  \ket{\psi_1}=G_0(E)T_1(E)\left(\U+\U^2\right)\ket{\psi_1}.
\end{equation}

\subsubsection{Basis set expansion approach}
In the following, we employ a basis set expansion approach 
to solve the Faddeev equation (\ref{eq:faddeev_equation_psi1-1}), which  
provides an extension of the momentum space Faddeev approach 
\cite{Gloeckle83} to 
systems of trapped atoms. Since, according to
Eq.~(\ref{eq:hamilton_HO_three_particles_jac}), the free Hamiltonian 
$H_0$  can be divided into a sum of three independent harmonic 
oscillators, we choose the basis of products 
\begin{equation}
\ket{\Phi_{KLM_L},\phi_{\kappa\lambda\mu_\lambda},\varphi_{klm_l}}
=\ket{\Phi_{KLM_L}}\ket{\phi_{\kappa\lambda\mu_\lambda}}
\ket{\varphi_{klm_l}}
\end{equation} 
of the energy states of the individual oscillators with a definite angular 
momentum, i.e.~$\ket{\Phi_{KLM_L}}$,  
$\ket{\phi_{\kappa\lambda\mu_\lambda}}$ and $\ket{\varphi_{klm_l}}$
are the oscillator energy states associated with the Jacobi coordinates 
$\vec R$, $\vec \rho$ and $\vec r$, respectively.

The kernel of the Faddeev equation (\ref{eq:faddeev_equation_psi1-1}) is  
diagonal in the non-interacting energy states $\ket{\Phi_{KLM_L}}$. The 
component $\ket{\psi_1}$ can thus be chosen in such a way that it 
factorises into a centre of mass part and a relative part, i.e.
\begin{equation}
  \label{eq:faddeev_equation_psi-factorise}
  \ket{\psi_1}=\ket{\Phi_{KLM_L}}\ket{\psi^\rel_1}.
\end{equation}
The relative part $\ket{\psirel}$ then satisfies a reduced Faddeev equation
at the shifted energy $\Erel=E-E_{KL}$:
\begin{equation}
  \label{eq:faddeev_equation_psi1KLM}
  \ket{\psirel}=G_0^\rel\left(\Erel\right)
  T_1^\rel\left(\Erel\right)\left[\U+\U^2\right]\ket{\psirel}.
\end{equation}
Here $G_0^\rel\left(\Erel\right)$ and $T_1^\rel\left(\Erel\right)$ can be 
interpreted in terms of a reduced non-interacting Green's function and a 
reduced $T$ matrix, respectively, which depend only on the Jacobi 
coordinates $\vec \rho$ and $\vec r$ describing the relative motion of the 
three atoms.

In order to solve Eq.~(\ref{eq:faddeev_equation_psi1KLM}),
the basis set expansion approach takes advantage of the 
convenient diagonal representation of the reduced non-interacting 
Green's function in terms of the chosen basis states:
\begin{align}
G_0^\rel(\Erel)=\sum_{\kappa,k\lambda,l=0}^\infty
\sum_{\mu_\lambda=-\lambda}^\lambda 
\sum_{m_l=-l}^l
\frac{\ket{\phi_{\kappa\lambda\mu_\lambda},\varphi_{klm_l}}
\bra{\phi_{\kappa\lambda\mu_\lambda},\varphi_{klm_l}}}
{\Erel-E_{\kappa\lambda}-E_{kl}}.
\label{eq:Green's_function_spectral}
\end{align}
Since the reduced $T$ matrix $T_1^\rel\left(\Erel\right)$ includes
just the interaction between the pair of atoms $(2,3)$, it is related 
to the $T$ matrix $\hat{T}(z)$ of the relative motion of this atom pair 
by the formula:
\begin{align}
\nonumber
\bra{\phi_{\kappa\lambda\mu_\lambda},\varphi_{klm_l}}
T_1^\mathrm{rel}(\Erel)
\ket{\phi_{\kappa'\lambda'\mu_\lambda'},\varphi_{k'l'm_l'}}=&
\bra{\varphi_{klm_l}}
\hat{T}(z)\ket{\varphi_{k'l'm_l'}}\\
&\times
\delta_{\kappa\kappa'}\delta_{\lambda\lambda'}
\delta_{\mu_\lambda\mu'_\lambda}.
\label{eq:relative_T_matrix}
\end{align}
Here $z=\Erel-E_{\kappa\lambda}$ accounts for the energy
of atom 1. The kernel of Eq.~(\ref{eq:faddeev_equation_psi1-1}) 
is thus completely determined by the solution of the two-body 
Lippmann-Schwinger equation in the presence of the trapping 
potential. 

The complete three-body energy state can be factorised in analogy to 
Eq.~(\ref{eq:faddeev_equation_psi-factorise}), i.e.
\begin{equation}
  \label{eq:psi-factorise}
  \ket{\psi}=\ket{\Phi_{KLM_L}}\ket{\psi_\rel}.
\end{equation}
Given the solution of Eq.~(\ref{eq:faddeev_equation_psi1KLM}),
the reduced Faddeev component $\ket{\psi_1^\mathrm{rel}}$ determines
$\ket{\psi_\mathrm{rel}}$ by the relationship:
\begin{equation}
  \label{eq:faddeev-decomposition}
  \ket{\psi_\rel}=\left(1+\U+\U^2\right)\ket{\psi^\rel_1}.
\end{equation}
Equations (\ref{eq:faddeev_equation_psi1KLM}), 
(\ref{eq:Green's_function_spectral}) and (\ref{eq:relative_T_matrix})
set up our general approach to the energy spectrum of three 
interacting atoms in a trap, while Eqs.~(\ref{eq:psi-factorise}) and
(\ref{eq:faddeev-decomposition}) yield the associated three-body 
energy states.

\subsection{Solution of the Faddeev equations in the separable potential 
  approach}  
\subsubsection{Faddeev equations in the separable potential approach}
We shall show in the following that the separable potential approach to the 
two-body $T$ matrix, in combination with the basis set expansion,
provides a practical scheme to exactly solve the Faddeev equation 
(\ref{eq:faddeev_equation_psi1KLM}) in the presence of a trapping potential.  
To this end, we apply Eqs.~(\ref{eq:relative_T_matrix}) and 
(\ref{eq:T-Matrix-separable-1}) to Eq.~(\ref{eq:faddeev_equation_psi1KLM}).
This yields:
\begin{align}
\nonumber
  \ket{\psirel}=&
  G_0^\rel\left(\Erel\right)
  \sum_{\kappa=0}^\infty\sum_{\lambda=0}^\infty
  \sum_{\mu_\lambda=-\lambda}^\lambda
  \ket{\phi_{\kappa\lambda\mu_\lambda},g}
  \tau\left(\Etri-E_{\kappa\lambda}\right)\\
  &\times
  \bra{\phi_{\kappa\lambda\mu_\lambda},g}
  \left[\U+\U^2\right]\ket{\psirel}.
  \label{eq:faddeev_equation_psi1KLM-1}
\end{align}
Equation (\ref{eq:faddeev_equation_psi1KLM-1}) reveals that $\ket{\psirel}$ 
is of the general form:
\begin{equation}
  \label{eq:faddeev_component_general_form}
  \ket{\psirel}=G_0^\rel\left(\Erel\right)\ket{f,g}.
\end{equation}
As only the unknown amplitude 
$f_{\kappa\lambda\mu_\lambda}=
\braket{\phi_{\kappa\lambda\mu_\lambda}}{f}$ needs to be determined,
the separable potential approach significantly simplifies  
the Faddeev equation (\ref{eq:faddeev_equation_psi1KLM}) by reducing its 
dimensionality from six to three. We shall show in the following that due to 
the spherical symmetry of the form factor $\ket g$ the number of dimensions 
reduces even further to only one provided that the three-body energy levels 
under consideration have $s$ wave symmetry.

\subsubsection{Basis set expansion}
In accordance with the basis set expansion approach, we consider  
the projected amplitude $f_{\kappa\lambda\mu_\lambda}$.
The {\em ansatz} of Eq.~(\ref{eq:faddeev_component_general_form}) for the 
solution of the Faddeev equation (\ref{eq:faddeev_equation_psi1KLM-1}) then 
determines the amplitude $f_{\kappa\lambda\mu_\lambda}$ by the matrix 
equation:
\begin{align}
  f_{\kappa\lambda\mu_\lambda}=\tau\left(\Erel-E_{\kappa\lambda}\right)
  \sum_{\kappa'=0}^\infty
  \sum_{\lambda'=0}^\infty
  \label{eq:faddeev_equation_psi1-coeff-3}
  \sum_{\mu_\lambda'=-\lambda'}^{\lambda'}
  \mathcal{K}_{\kappa\lambda\mu_\lambda,\kappa'\lambda'\mu_\lambda'}
  \left(\Erel\right) f_{\kappa'\lambda'\mu_\lambda'}.
\end{align}
In accordance with Eq.~(\ref{eq:Green's_function_spectral}),
the reduced kernel matrix associated with this equation for 
$f_{\kappa\lambda\mu_\lambda}$ is given by the formula:
\begin{align}
  \label{eq:faddeev_kernel_op}
  \mathcal{K}_{\kappa\lambda\mu_\lambda,\kappa'\lambda'\mu_\lambda'}
  \left(\Erel\right)=
  \bra{\phi_{\kappa\lambda\mu_\lambda},g}
  \left(\U+\U^2\right)G_0^\rel\left(\Erel\right)
  \ket{\phi_{\kappa'\lambda'\mu_\lambda'},g}.
\end{align} 
The complete kernel also involves the function 
$\tau\left(\Erel-E_{\kappa\lambda}\right)$ which we have discussed in detail 
in Appendix \ref{SubSecApp:Scattering_in_free_space}.
Inserting the spectral decomposition of the Green's function of three 
non-interacting trapped atoms of Eq.~(\ref{eq:Green's_function_spectral}) 
then determines the reduced kernel matrix to be:
\begin{align}
  \nonumber
  \mathcal{K}_{\kappa\lambda\mu_\lambda,\kappa'\lambda'\mu_\lambda'}
  =&\sum_{k,k'=0}^\infty \sum_{l,l'=0}^\infty
  \sum_{m_l=-l}^l\sum_{m_l'=-l'}^{l'}
  \frac{\braket{g}{\varphi_{klm_l}}\braket{\varphi_{k'l'm_l'}}{g}}
       {\Erel-E_{\kappa'\lambda'}-E_{k'l'}} \\    
  &\times
  \bra{\phi_{\kappa\lambda\mu_\lambda},\varphi_{klm_l}}
  \left(\U+\U^2\right)
  \ket{\phi_{\kappa'\lambda'\mu_\lambda'},\varphi_{k'l'm_l'}}.
  \label{eq:faddeev_kernel}
\end{align}

\subsubsection{Symmetry considerations}
The spherical symmetry of the form factor $\ket{g}$
of Eq.~(\ref{eq:formfactor_mom}) implies that only the spherically symmetric 
basis states $\ket{\varphi_k}\equiv\ket{\varphi_{k00}}$ contribute to the 
summation in Eq.~(\ref{eq:faddeev_kernel}), i.e.~$l=l'=0$ and $m_l=m_l'=0$. 
As, moreover, the total angular momentum operator 
$\vec{\mathcal L}=\vec\lambda+\vec l$ associated with 
the three-body state $\ket{\psi_\rel}$ commutes with the permutation operator
$\U$, the off-diagonal elements $\lambda\neq\lambda'$ and $\mu\neq\mu_\lambda'$
of the kernel vanish, i.e.~the kernel does not couple solutions of
different angular momenta $\vec{\lambda}$. Furthermore, the Hamiltonian
$H$ commutes with $\U$ which implies that the matrix element of
$\left(\U+\U^2\right)$ in the reduced kernel matrix of 
Eq.~(\ref{eq:faddeev_kernel}) 
is non-zero only if
\begin{equation}
  \label{eq:energy_conservation}
  \kappa+k=\kappa'+k'.
\end{equation}
This is a consequence of energy conservation.

We shall focus in the following on those energy states of three trapped 
interacting atoms that correlate adiabatically, in the limit of zero trap 
frequency, with the three-body $s$ wave Efimov states.
We thus restrict the discussion to three-body states with zero total
angular momentum $\mathcal L=0$. In this case all angular momentum
quantum numbers are zero and may be omitted. This implies that the kernel 
matrix in Eq.~(\ref{eq:faddeev_equation_psi1-coeff-3}) reduces to
\begin{equation}
  \label{eq:faddeev_kernel-1}
  \mathcal{K}_{\kappa\kappa'}\left(\Erel\right)=\sum_{k,k'=0}^\infty
  \frac{g_k \bra{\phi_\kappa,\varphi_k}\left(\U+\U^2\right) 
    \ket{\phi_{\kappa'},\varphi_{k'}}  g_{k'}}
  {\Erel-E_\kappa-E_k},
\end{equation}
and the amplitude $f_\kappa=\braket{\phi_\kappa}{f}$ satisfies the matrix 
equation:
\begin{equation}
  \label{eq:faddeev_equation_psi1-coeff-4}
  f_\kappa=\tau\left(\Erel-E_\kappa\right)\sum_{\kappa'=0}^\infty
  \mathcal{K}_{\kappa\kappa'}\left(\Erel\right)f_{\kappa'}.
\end{equation}
In accordance with the {\em ansatz} of 
Eq.~(\ref{eq:faddeev_component_general_form}), the
first Faddeev component is then given in terms of the amplitude $f_\kappa$
and the form factor $g_k$ by the formula:
\begin{eqnarray}
  \label{eq:faddeev_equation_psi1-coeff-5}
  \braket{\phi_\kappa,\varphi_k}
  {\psi_{1}^\rel}= \frac{f_{\kappa} g_k}{\Erel-E_{\kappa}-E_{k}}.
\end{eqnarray}

\subsection{Determination of the kernel matrix \label{sec:kernel}}
A main difficulty in the numerical determination of the amplitude 
$f_\kappa$ from Eq.~(\ref{eq:faddeev_equation_psi1-coeff-4}) consists in 
calculating the reduced kernel matrix 
$\mathcal{K}_{\kappa\kappa'}\left(\Erel\right)$
for the variety of trap frequencies studied in this paper.
While in tight atom traps ($\nu_\mathrm{ho}>100\,$kHz) the discrete nature 
of the energy levels is most significant, the opposite regime of low 
trap frequencies ($\nu_\mathrm{ho}<1\,$kHz) involves a large range of 
vibrational quantum numbers $\kappa$ leading to a continuum of modes in
the limit $\nu_\mathrm{ho}\to 0$. To obtain a stable scheme for the 
determination of the reduced kernel matrix in the 
limits of both high and low trap 
frequencies, we have performed separate treatments of the regimes of low and 
high vibrational quantum numbers $\kappa$. Since these aspects of the studies 
of trapped systems of three interacting atoms differ significantly from the
known techniques to solve the Faddeev equations in free space 
\cite{Gloeckle83}, we shall outline in detail the numerical procedure we have 
applied. 

\subsubsection{The kernel matrix at low vibrational excitations}
In the limit of small $\kappa$ the determination of the reduced kernel matrix
$\mathcal{K}_{\kappa\kappa'}$ consists mainly of the calculation
of the matrix elements 
\begin{equation}
  C_{\kappa k,\kappa' k'}=\bra{\phi_\kappa,\varphi_k}\left[\U+\U^2\right]
  \ket{\phi_{\kappa'},\varphi_{k'}}.
\end{equation} 
We perform this calculation 
in the configuration space representation. To this end, it is convenient 
to introduce the auxiliary function:
\begin{equation}
  \label{eq:wavefunction_HO_one_particle_generic}
  \psi_k(\beta; x)=\sqrt{\frac{\beta^{3/2}}{2\pi} 
    \frac{\Gamma(k+1)}{\Gamma(k+\frac32)}}
  \ e^{-x/2} \ \LagP{k}(x).
\end{equation}
This function is related to the spherically symmetric harmonic oscillator 
states (cf.~Appendix \ref{SubSecApp:Two_body_energy_levels}) 
associated with the Jacobi coordinates $\vec{\rho}$ and $\vec{r}$ by 
the formulae
\begin{align}
  \phi_\kappa(\rho)=&\psi_\kappa(\beta_\rho; \beta_\rho \rho^2),\\
  \varphi_k(r)=&\psi_k(\beta_r; \beta_r r^2),
\end{align}
respectively.
Here the parameters $\beta_\rho=\frac{2}{3}m\omega_\mathrm{ho}/\hbar$ and 
$\beta_r=\frac{1}{2}m\omega_\mathrm{ho}/\hbar$ account for the different 
masses associated with the individual harmonic oscillator contributions to 
the three-body Hamiltonian (\ref{eq:hamilton_HO_three_particles_jac}).
The matrix element involving the permutation operator can then be 
represented by:
\begin{align}
  \nonumber
  C_{\kappa k,\kappa' k'}=&
  \int d^3\rho\ d^3 r\
  \psi_\kappa(\beta_\rho;\beta_\rho\rho^2)
  \psi_k(\beta_r;\beta_r r^2)\\
  \nonumber
  &\times
  \left[
    \psi_{\kappa'}(\beta_\rho;\beta_\rho\rho'^2)
    \psi_{k'}(\beta_r;\beta_r r'^2)
    \right.\\
    &+
    \left.
    \psi_{\kappa'}(\beta_\rho;\beta_\rho\rho''^2)
    \psi_{k'}(\beta_r;\beta_r r''^2)
    \right].
  \label{eq:Uelement-0}
\end{align}
Here the permutation operators $\U$ and $\U^2$ transform the
coordinates $\vec\rho$ and $\vec{r}$ into primed and double primed 
coordinates $\textstyle{{\vec{\rho}'}\choose{\vec{r}'}}=
\md\textstyle{{\vec\rho}\choose{\vec r}}$ and
$\textstyle{{\vec{\rho}''}\choose{\vec{r}''}}=
\md^2\textstyle{{\vec\rho}\choose{\vec r}}$, respectively, in accordance 
with the transformation matrix $\md$ of 
Eq.~(\ref{eq:md-definition}). The transformation with $\md$ yields:
\begin{align}
  \beta_\rho\rho'^2=&\frac14\beta_\rho\rho^2+\frac34\beta_r
  r^2-\sqrt{\frac34\beta_\rho\beta_r}\ \vec\rho\cdot\vec r,\\
  \beta_r r'^2=&\frac34\beta_\rho\rho^2+\frac14\beta_r
  r^2+\sqrt{\frac34\beta_\rho\beta_r}\ \vec\rho\cdot\vec r.
\end{align}
This implies the relationship:
\begin{equation}
\beta_\rho\rho'^2+\beta_r r'^2=\beta_\rho\rho^2+\beta_r r^2. 
\end{equation}
Similar relations hold for the double primed coordinates with
the reversed signs in front of the square roots. The  
integrand on the right hand side of Eq.~(\ref{eq:Uelement-0}), therefore, 
depends only on the variables $\rho$ and $r$ in addition to the   
variable $x=\vec{\rho}\cdot\vec{r}/(\rho r)$, which involves the  
angle between the coordinates $\vec{\rho}$ and $\vec{r}$. The 
integration over the remaining three variables is readily performed 
and leads to the formula:
\begin{align}
  \nonumber
  C_{\kappa k,\kappa' k'}=&
  4(2\pi)^2\int_0^\infty\rho^2 d\rho\ \int_0^\infty r^2 d r\
  \psi_\kappa(\beta_\rho;\beta_\rho\rho^2)
    \psi_k(\beta_r;\beta_r r^2)\\
    \nonumber
    &\times
    \int_{-1}^{1} d x \
    \psi_{\kappa'}\left(\beta_\rho;\frac14\beta_\rho\rho^2+\frac34\beta_r
    r^2-\sqrt{\frac34\beta_\rho\beta_r}\rho rx\right)\\
    &\times
    \psi_{k'}\left(\beta_r;\frac34\beta_\rho\rho^2+\frac14\beta_r
    r^2+\sqrt{\frac34\beta_\rho\beta_r}\rho rx\right).
    \label{eq:Uelement-1}
\end{align}
This formula can be further evaluated by changing the variables to 
$u=\beta_\rho\rho^2$ and $v=\beta_r r^2$ and using the explicit form of 
the harmonic oscillator wave functions of
Eq.~(\ref{eq:wavefunction_HO_one_particle_generic}). This evaluation yields:
\begin{align}
  C_{\kappa k,\kappa' k'}=
  \sqrt{\frac{\Gamma(\kappa+1)}{\Gamma(\kappa+\frac32)}
    \frac{\Gamma(k+1)}{\Gamma(k+\frac32)}
    \frac{\Gamma(\kappa'+1)}{\Gamma(\kappa'+\frac32)}
    \frac{\Gamma(k'+1)}{\Gamma(k'+\frac32)}}\ I_{\kappa k,\kappa' k'}.
\label{eq:U_kappakappapkkp}
\end{align}
The coefficients $I_{\kappa k,\kappa' k'}$ that involve the associated 
Laguerre polynomials read:
\begin{align}
  \nonumber
  I_{\kappa k,\kappa' k'}=&
  \int_0^\infty d u\ u^{\frac12} e^{-u} \LagP{\kappa}(u)
  \int_0^\infty d v\ v^{\frac12} e^{-v} \LagP{k}(v)\\
  &\times
  \int_{-1}^{1} d x\ P_{\kappa' k'}(u,v,x).
   \label{eq:I_kappakappapkkp}
\end{align}
Here the function
\begin{align}
  \nonumber
  P_{\kappa' k'}(u,v,x)=&
    \LagP{\kappa'}\left(\frac14 u+\frac34 v-
    \sqrt{\frac34 uv} x\right)\\
  &\times 
  \LagP{k'}\left(\frac34 u+\frac14 v+
  \sqrt{\frac34 uv} x\right)
  \label{eq:Puvx}
\end{align}
is a polynomial of degree $\kappa'+k'$ in the variable $x$. 
These derivations reveal that the matrix element in 
Eq.~(\ref{eq:U_kappakappapkkp}) is independent not only of the inter-atomic 
interaction potential but also of the frequency of the atom trap. The trap 
frequency enters Eq.~(\ref{eq:faddeev_equation_psi1-coeff-4}) through the 
projections of the form factor onto the basis states and through the energy 
denominator.

To further evaluate Eq.~(\ref{eq:I_kappakappapkkp}), we represent the function 
$P_{\kappa' k'}(u,v,x)$ by the sum:
\begin{equation}
  \label{eq:Puv_m}
  P_{\kappa' k'}(u,v,x)=\sum_{\syc=0}^{\kappa'+k'}
  P_{\kappa'k'}^{\syc}(u,v) x^{\syc}.
\end{equation}
Here the coefficients $P_{\kappa'k'}^{\syc}(u,v)$ depend on the variables $u$ 
and $v$. Equation (\ref{eq:Puv_m}) then allows us to perform the integration 
over the variable $x$ in Eq.~(\ref{eq:I_kappakappapkkp}). 
Only the even powers $x^{\syc}$ of the variable $x$ contribute to this 
integral, while all terms involving the odd powers vanish. When $\syc$ is 
even, however, it turns out that the coefficients $P_{\kappa'k'}^{\syc}(u,v)$ 
themselves are bivariate polynomials in the variables $u$ and $v$ and can also 
be expanded in powers of $u$ and $v$ with expansion coefficients 
$P_{\kappa'k'}^{\syc\sya\syb}$. This expansion thus yields:
\begin{equation}
  \label{eq:Puv_mab}
  P_{\kappa'k'}^{\syc}(u,v)=\sum_{\sya=0}^{\kappa'+k'}
  \sum_{\syb=0}^{\kappa'+k'-\sya} P_{\kappa'k'}^{\syc\sya\syb}
  u^{\sya}v^{\syb}.
\end{equation}

The representation of the expansion coefficients 
$P_{\kappa'k'}^{\syc}(u,v)$ in Eq.~(\ref{eq:Puv_mab}) 
in terms of a polynomial allows us to 
take advantage of the general formula
\begin{align}
  \int_0^\infty d u\ u^\frac{1}{2} e^{-u} \LagP{\kappa}(u) u^\sya
  =\left\{\begin{array}{c@{\,:\,}r}
  0 & 0\le \sya <\kappa\\
  (-1)^\kappa \Gamma(\sya+\frac32)
  \binom{\sya}{\kappa} & \qquad \kappa\le \sya
  \end{array}\right.
  \label{eq:LagInt}
\end{align}
to perform the remaining integrations in Eq.~(\ref{eq:I_kappakappapkkp})
over the variables $u$ and $v$. Here $\binom{\sya}{\kappa}$ is a 
combinatorial. Equation (\ref{eq:LagInt}) thus determines the coefficients 
$I_{\kappa k,\kappa' k'}$ of Eq.~(\ref{eq:I_kappakappapkkp}) to be:
\begin{align}
  \nonumber
  I_{\kappa k,\kappa' k'}=&(-1)^{\kappa'+k'}
    \sum^{\kappa'+k'}_{{\syc=0}\atop{(\syc\
        \mathrm{even})}}\!\!
    \frac{2}{\syc+1}
    \sum_{\sya=\kappa}^{\kappa'+k'}
    \Gamma\left(\sya+\frac32\right)\binom{\sya}{\kappa}\\
  &\times
  \sum_{\syb=k}^{\kappa'+k'-\sya}
  \Gamma\left(\syb+\frac32\right) \binom{\syb}{k}
   P_{\kappa'k'}^{\syc\sya\syb}.
   \label{eq:I_kappakappapkkp_1}
\end{align}

The limits of the different sums in Eq.~(\ref{eq:I_kappakappapkkp_1})
lead to a further simplification in the determination of 
$I_{\kappa k,\kappa' k'}$ as follows: The summation over the index $\syb$ 
is limited by the condition $\kappa'+k'-\sya\ge k$, which, together with the 
condition $\sya\ge \kappa$ for the index $\sya$, implies the inequality 
$\kappa'+k'\ge\kappa+k$. Similar considerations show that also the 
inequality $\kappa'+k'\le\kappa+k$ is fulfilled, which, in summary, leads to 
the restriction $\kappa+k=\kappa'+k'$. These explicit derivations simply 
recover Eq.~(\ref{eq:energy_conservation}), which we have obtained 
independently from general symmetry considerations. Consequently, the 
summations over $\sya$ and $\syb$ in Eq.~(\ref{eq:I_kappakappapkkp_1}) 
reduce to a single term determined by $\sya=\kappa$ and $\syb=k$. 
This gives the coefficient $I_{\kappa k,\kappa' k'}$ to be:
\begin{align}
  I_{\kappa k,\kappa' k'}=(-1)^{\kappa+k}\
  \Gamma\left(\kappa+\frac32\right) \Gamma\left(k+\frac32\right)
  \sum^{\kappa+k}_{{\syc=0}\atop{(\syc\
      \mathrm{even})}}\!\! \frac{2}{\syc+1}
   P_{\kappa'k'}^{\syc\kappa k}. 
   \label{eq:I_kappakappapkkp_2}
\end{align}

The remaining, as yet, undetermined expansion coefficients
$P_{\kappa'k'}^{\syc\kappa k}$ of Eq.~(\ref{eq:Puv_mab})
can be obtained from the explicit form of the associated Laguerre 
polynomials, 
i.e.~$\LagP{\kappa}(x)=\sum_{\syd=0}^\kappa c_{\kappa\syd} x^\syd$,
and their expansion coefficients:
\begin{equation}
  \label{eq:LagCoeff}
  c_{\kappa \syd}=(-1)^\syd\frac1{\syd!}\binom{\kappa+\frac12}{\kappa-\syd}.
\end{equation}
To this end, we consider the function $P_{\kappa'k'}^{\syc}(u,v)$ of 
Eq.~(\ref{eq:Puv_m}), whose explicit form can be determined from 
Eq.~(\ref{eq:Puvx}) to be: 
\begin{align}
\nonumber
  P_{\kappa'k'}^{\syc}(u,v)=&\left(\frac34\right)^{\frac{\syc}2}
  (uv)^{\frac{\syc}2}
  \sum_{\syd=0}^{\kappa'} c_{\kappa'\syd}\sum_{\sye=0}^{k'}c_{k'\sye}
  \sum_{\syf=\max(0,\syc-\sye)}^{\min(\syd,\syc)} (-1)^{\syf}
  \binom{\syd}{\syf}\\ 
  &\times
  \binom{\sye}{\syc-\syf}
  \left(\frac14  u+\frac34 v\right)^{\syd-\syf}
  \left(\frac34  u+\frac14 v\right)^{\sye+\syf-\syc}.
  \label{eq:Puv_m_explicit}
\end{align}
Equations (\ref{eq:Puv_mab}) and (\ref{eq:Puv_m_explicit}) then give the 
remaining expansion coefficients $P_{\kappa'k'}^{\syc\kappa k}$ to be:
\begin{align}
\nonumber
  P_{\kappa'k'}^{\syc\kappa k}=&
  4^{\frac{\syc}2-\kappa-k}
  c_{\kappa'\kappa'}c_{k'k'}
  \!\!\!
  \sum_{\syf=\max(0,\syc-k')}^{\min(\kappa',\syc)} 
  \!\!\!\!\!\! (-1)^{\syf}
  \binom{\kappa'}{\syf}
  \binom{k'}{\syc-\syf}\\
  &\times
  \sum_{\syg=\max(0,\kappa'-\kappa-
    \syf+\frac{\syc}2)}^{\min(\kappa'-\syf,k-\frac{\syc}2)}
  \binom{\kappa'-\syf}{\syg}
  \binom{k'+\syf-\syc}{k-\frac{\syc}2-\syg}
  \ 3^{\kappa-\kappa'+\syf+2\syg}. 
  \label{eq:Puv_mab_explicit_1}
\end{align}

In order to summarise our results in a concise form, we introduce the 
coefficients
\begin{equation}
  \label{eq:Puv_mab_explicit_numerical}
  \tilde{P}_{\kappa'k'}^{\syc\kappa k}=
  \frac{P_{\kappa'k'}^{\syc\kappa k}}{c_{\kappa'\kappa'}c_{k'k'}}=
  (-1)^{\kappa+k}\ 
  \Gamma(\kappa'+1)\Gamma(k'+1)
  P_{\kappa'k'}^{\syc\kappa k},
\end{equation}
which are obtained simply by dividing $P_{\kappa'k'}^{\syc\kappa k}$ 
of Eq.~(\ref{eq:Puv_mab_explicit_1}) by the expansion coefficients
$c_{\kappa'\kappa'}=(-1)^{\kappa'}/\Gamma(\kappa'+1)$ and 
$c_{k'k'}=(-1)^{k'}/\Gamma(k'+1)$ 
of the associated Laguerre polynomials.
An analysis of its summation limits shows that the sum over $\syg$ 
in Eq.~(\ref{eq:Puv_mab_explicit_1}) vanishes unless the conditions
$\syc\le 2\kappa$ and $\syc\le 2k$ are fulfilled. In accordance with 
Eq.~(\ref{eq:I_kappakappapkkp_2}), we thus obtain:
\begin{equation}
  \label{eq:I_kappakappapkkp_result}
  I_{\kappa k,\kappa' k'}=
  \frac{\Gamma\left(\kappa+\frac32\right)\Gamma\left(k+\frac32\right)}
  {\Gamma\left(\kappa'+1\right)\Gamma\left(k'+1\right)}
  \sum^{\min(\kappa+k,2\kappa,2k)}_{{\syc=0}\atop{(\syc\ \mathrm{even})}}
  \frac{2}{\syc+1} \tilde{P}_{\kappa'k'}^{\syc\kappa k}.
\end{equation}
The complete matrix element of Eq.~(\ref{eq:Uelement-0}) is then 
determined by the formula:
\begin{align}
  \nonumber
  C_{\kappa k,\kappa' k'}=&
  \sqrt{\frac{\Gamma(\kappa+1)}{\Gamma(\kappa'+1)}
    \frac{\Gamma(\kappa+\frac32)}{\Gamma(\kappa'+\frac32)}
    \frac{\Gamma(k+1)}{\Gamma(k'+1)}
    \frac{\Gamma(k+\frac32)}{\Gamma(k'+\frac32)}}\\
  &\times
  \sum^{\min(\kappa+k,2\kappa,2k)}_{{\syc=0}\atop{(\syc\ \mathrm{even})}}
  \!\!
  \frac{2}{\syc+1} \tilde{P}_{\kappa'k'}^{\syc\kappa k}.
  \label{eq:U_kappakappapkkp-1}
\end{align}

Since the triple summation over $\syf$, $\syg$ and $\syc$ in 
Eqs.~(\ref{eq:Puv_mab_explicit_1}) and (\ref{eq:U_kappakappapkkp-1})
contains only integer numbers, the calculation of the matrix element 
$C_{\kappa k,\kappa' k'}$ is, in principle, straightforward. As these
numbers can, however, be very large in magnitude and have alternating 
signs, we have employed a computer algebra system to carry out the sum in
multi-precision integer arithmetic. This proved practical for the indices
$0\le\kappa,\kappa'\lesssim 40$.

\subsubsection{The kernel matrix in the limit of high vibrational excitations}
The numerical determination of the matrix element $C_{\kappa k,\kappa' k'}$ 
using Eq.~(\ref{eq:U_kappakappapkkp-1}) becomes impractical in the limit of 
large indices $\kappa$ and $\kappa'$. We shall, therefore, provide a scheme 
to directly determine the reduced kernel matrix for these higher vibrational 
quantum numbers in terms of its asymptotic form in the continuum limit. 
Starting from Eq.~(\ref{eq:faddeev_kernel_op}) evaluated at zero angular 
momenta $\lambda=0$ and $\lambda'=0$, the exact reduced kernel matrix is 
given by the formula:
\begin{align}
  \nonumber
  \mathcal{K}_{\kappa\kappa'} \left(\Erel\right)=&
  \bra{\phi_{\kappa},g}
  \left(\U+\U^2\right)G_0^\rel\left(\Erel\right)
  \ket{\phi_{\kappa'},g}\\
  =&\sum_{k=0}^\infty\frac{\bra{\phi_{\kappa},g}\left(\U+\U^2\right)
  \ket{\phi_{\kappa'},\phi_{k}}\braket{\phi_{k}}{g}}{\Erel-E_{\kappa'}-E_k}.
  \label{eq:faddeev_kernel_op-1}
\end{align}
It turns out that, due to the permutation operators, the right hand side of 
Eq.~(\ref{eq:faddeev_kernel_op-1}) depends just on the oscillator wave 
functions $\phi_{\kappa}(\rho)$, $\phi_{\kappa'}(\rho)$ and $\phi_k(r)$
in a limited range of radii $\rho$ and $r$ set by the width of the 
form factor $g(r)$. The typical length scale associated with the width of 
$g(r)$ is set by the range parameter $\sigma$ of 
Eq.~(\ref{eq:range_parameter_sigma}) and is thus determined by the van der 
Waals length $l_\mathrm{vdW}$. As the van der Waals length is
typically much smaller than the harmonic oscillator length 
$a_\mathrm{ho}=\sqrt{\hbar/(m\omega_\mathrm{ho})}$, the trapping potential 
is flat within the relevant range of radii $\rho,r\lesssim \sigma$ and the 
potential energy is much smaller than the mean kinetic energy of the highly 
excited oscillator states. We can, therefore, replace in 
Eq.~(\ref{eq:faddeev_kernel_op-1}) the Green's function 
$G_0^\rel\left(\Erel\right)$ of the trapping potential by its counterpart
$G_0^{\mathrm{free}}\left(\Erel\right)$ in free space and perform the 
continuum limit of the oscillator states $\ket{\phi_\kappa}$ and 
$\ket{\phi_{\kappa'}}$.

Given that the radius $\rho$ is limited by the condition $\rho\lesssim \sigma$,
in the continuum limit the harmonic oscillator energy wave functions 
$\phi_{\kappa\lambda\mu_\lambda}(\vec{\rho})$ approach, up to a 
normalisation constant, the partial waves $\ket{q\lambda\mu_\lambda}$,
i.e.~the improper energy states of the free space Hamiltonian
\begin{equation}
  H_\rho=-\frac{\hbar^2}{2(\frac23m)}\nabla_{\vec\rho}^2
\end{equation}
with a definite angular momentum.

These partial waves thus satisfy the Schr\"odinger equation 
$H_\rho\ket{q\lambda\mu_\lambda}=E_q \ket{q\lambda\mu_\lambda}$, 
associated with the kinetic energy $E_q=q^2/[2(\frac23m)]$, and are related 
to the plane waves $\ket{\vec q'}$ by 
\begin{equation}
\braket{\vec q'}{q\lambda\mu_\lambda}=(-i)^\lambda\ q^{-2}\delta(q-q') 
Y_\lambda^{\mu_\lambda}(\vec{q}'/q'), 
\end{equation}
where $Y_\lambda^{\mu_\lambda}$ is a spherical harmonic. 

To determine the asymptotic form of the reduced kernel matrix in the limit of 
high vibrational excitations, we insert into Eq.~(\ref{eq:faddeev_kernel_op-1})
two complete sets of improper states $\ket{q\lambda\mu_\lambda}$ and 
$\ket{q'\lambda'\mu_\lambda'}$. This yields:
\begin{align}
  \nonumber
  \mathcal{K}_{\kappa\kappa'} 
  \left(\Erel\right)\underset{\kappa,\kappa'\to\infty}{\sim}&
  \int_0^\infty\!\!\!d q\,q^2 \braket{\phi_{\kappa}}{q00}\!
  \int_0^\infty\!\!\!d q'\,q'^2 \braket{q'00}{\phi_{\kappa'}}\\
  &\times
  \bra{q00,g}
  \left(\U+\U^2\right)G_0^\mathrm{free}\left(\Erel\right)
  \ket{q'00,g}.
  \label{eq:faddeev_kernel_op-2}
\end{align}
A simple calculation then shows that for high vibrational quantum numbers
$\kappa$ the function $q^2\braket{\phi_{\kappa}}{q00}$ is sharply peaked about 
the central momentum $\bar{q}$ at the matched energies $E_\kappa=E_q$. 
This implies:
\begin{equation}
  \label{eq:Eq_is_Ekappa}
  \bar{q}=\sqrt{\frac43
    m\hbar\omega_\mathrm{ho}\left(2\kappa+\frac32\right)}.
\end{equation}
A similar relation holds for the central momentum $\bar{q}'$ associated with 
the function $q'^2\braket{q'00}{\phi_{\kappa'}}$.
We may, therefore, evaluate the slowly varying matrix element  
$\bra{q00,g}\left(\U+\U^2\right)G_0^\mathrm{free}\left(\Erel\right)
\ket{q'00,g}$
in Eq.~(\ref{eq:faddeev_kernel_op-2}) at  $q=\bar{q}$ and 
$q'=\bar{q}'$. The remaining integration over $q$ then yields
\begin{align}
  \int_0^\infty\!\!\!d q\,q^2 \braket{\phi_{\kappa}}{q00}=
  2\left(\frac{2m\hbar\omega_\mathrm{ho}}{3}\right)^{3/4}
  \sqrt{\frac{\Gamma(\kappa+\frac32)}{\Gamma(\kappa+1)}},
\end{align}
and the integration over $q'$ can be performed in an analogous way.
Using the spectral decomposition of the free space Green's function in terms
of plane wave momentum states and the explicit Gaussian expression for 
the form factor in Eq.~(\ref{eq:formfactor_mom}), the remaining matrix element 
$\bra{\bar{q}00,g}\left(\U+\U^2\right)G_0^\mathrm{free}
\left(\Erel\right)\ket{\bar{q}'00,g}$   
can be determined analytically. In the limit of high vibrational excitations 
$\kappa$ and $\kappa'$ the reduced kernel matrix is then given by the formula:
\begin{align}
  \nonumber
    \mathcal{K}_{\kappa\kappa'}\left(\Erel\right)
    \underset{\kappa,\kappa'\to\infty}{\sim}&
    \frac{16}{\sqrt\pi}
    \sqrt{\frac{\Gamma(\kappa+\frac32)}{\Gamma(\kappa+1)}
      \frac{\Gamma(\kappa'+\frac32)}{\Gamma(\kappa'+1)}}
    \left(\frac23 \frac{m\omega_\mathrm{ho}\sigma^2}\hbar \right)^{3/2}\\
    \nonumber
    &\times
    \frac{m}{\bar{q}\bar{q}'}\
    \exp\left(\left[\frac38(\bar{q}^2+\bar{q}'^2)-m\Erel\right]
    \sigma^2/\hbar^2\right)\\  
    \nonumber
    &\times
    \Bigg\{
    \Ei\left[\left(m\Erel-(\bar{q}^2+\bar{q}'^2-\bar{q}\bar{q}')\right)
      \sigma^2/\hbar^2\right]
    \\
    & -
    \Ei\left[\left(m\Erel-(\bar{q}^2+\bar{q}'^2+\bar{q}\bar{q}')\right) 
      \sigma^2/\hbar^2\right]
    \Bigg\}.
    \label{eq:faddeev_kernel-4}
\end{align}
Here $\Ei$ denotes the exponential integral
$\Ei(x)=\int_{-\infty}^x d t\ t^{-1} e^t$.

\subsection{Three-body energy wave functions}
\subsubsection{Basis set expansion of a three-body energy state}

Once the reduced kernel matrix $\mathcal{K}_{\kappa\kappa'}(\Erel)$ 
has been calculated, the three-body energies $\Erel$ and amplitudes 
$f_\kappa=\braket{\phi_\kappa}{f}$ 
can be obtained from the numerical solution of 
Eq.~(\ref{eq:faddeev_equation_psi1-coeff-4}). According to
Eqs.~(\ref{eq:faddeev-decomposition}) and 
(\ref{eq:faddeev_component_general_form}), each solution completely determines 
its associated three-body energy state by the formula:
\begin{equation}
  \label{eq:fullwavefunction}
  \ket{\psi_\rel}=\left(\Eins+\U+\U^2\right)
  G_0^\rel\left(\Erel\right)\ket{f,g}.
\end{equation}
The numerical determination of the complete state $\ket{\psi_\rel}$ requires
an expansion of Eq.~(\ref{eq:fullwavefunction}) into harmonic oscillator 
states. Since, throughout this paper, we focus on three-body $s$ wave states, 
a suitable basis set for the expansion is provided by the the three-body 
oscillator energy states with zero total angular momentum
$\vec{\mathcal L}=\vec\lambda+\vec l$. The total angular momentum quantum 
number of a three-body $s$ wave state is thus given by $\mathcal L=0$, and 
$\mathcal{M}_\mathcal{L}=0$ is its associated orientation quantum number. 
The $s$ wave basis states are then related to the oscillator energy states 
$\ket{\phi_{\kappa\lambda\mu_\lambda}}$ and $\ket{\varphi_{k l m_l}}$ 
with $\lambda=l$ and $\mu_\lambda=-m_l$ by the formula: 
\begin{equation}
  \label{eq:LM_basis}
  \ket{\phi_{\kappa}, \varphi_k;\mathcal L\!=\!0,
    \mathcal{M}_\mathcal{L}\!=\!0,l,l}=
  \!\!\sum_{m_l=-l}^l \frac{(-1)^{l-m_l}}{\sqrt{2l+1}}
  \ket{\phi_{\kappa lm_l},\varphi_{kl-m_l}}.
\end{equation}
Here ${(-1)^{l-m_l}}/{\sqrt{2l+1}}$ is a Clebsch-Gordan
coefficient. Although the permutation operator $\U$
commutes with the total three-body angular momentum 
$\vec{\mathcal L}$, it does not individually commute with the partial 
angular momenta $\vec\lambda$ and $\vec l$. Despite the fact that
$\ket f$ and $\ket g$ contain only contributions from the spherically 
symmetric harmonic oscillator basis states $\ket{\phi_{\kappa}}$ and 
$\ket{\varphi_{k}}$, respectively, the basis set expansion of the complete 
state $\ket{\psi_\rel}$ thus involves oscillator states 
$\ket{\phi_{\kappa\lambda\mu_\lambda}}$ and $\ket{\varphi_{k l m_l}}$ with 
all $\lambda,l=0,1,2,\dots$ values. This leads to the difficulty that the 
matrix elements of Eq.~(\ref{eq:Uelement-0}) need to be calculated also 
between these states. While this effort will generally be inevitable, we 
shall show that, for the purposes of this paper, it can be avoided. 
We utilise the fact that the 
determination of matrix elements, in the three-body energy states, of those 
quantities that commute with all three operators $\U$, $\vec \lambda$, 
and $\vec l$ involve just the spherically symmetric oscillator states 
$\ket{\phi_{\kappa}}$ and $\ket{\varphi_{k}}$. This is most evident for the 
normalisation constant and the orthogonality relation of the three-body 
energy states.

\subsubsection{Orthogonality and normalisation}


To derive the normalisation constant of the fully symmetrised three-body 
energy states as well as their orthogonality relation, we consider a pair of 
three-body states $\ket{\psi_\rel}$ and $\ket{\widetilde{\psi}_\rel}$ with the 
associated energies $\Erel$ and $\wErel$, respectively. Starting from
Eq.~(\ref{eq:fullwavefunction}) and using the identity $\U^3=1$, the overlap 
between these states is given, in terms of their first Faddeev components, by
the matrix element:
\begin{equation}
  \label{eq:trimer_orthogonality}
  \braket{\widetilde{\psi}_\rel}{\psi_\rel}=3
  \bra{\widetilde{\psi}^\rel_1}\left(\Eins+\U+\U^2\right)\ket{\psi^\rel_1}.
\end{equation}
Since both $\ket{\psi^\rel_1}$ and $\ket{\widetilde{\psi}^\rel_1}$
can be expanded in terms of the spherically symmetric oscillator states
$\ket{\phi_{\kappa}}$ and $\ket{\varphi_{k}}$, we obtain the formula:
\begin{align}
  \nonumber
  \braket{\widetilde{\psi}_\rel}{\psi_\rel}=&3\sum_{\kappa,k=0}^\infty 
  \braket{\widetilde{\psi}^\rel_1}{\phi_\kappa, \varphi_k}
  \Bigg[
    \braket{\phi_\kappa, \varphi_k}{\psi^\rel_1}\\
    \label{eq:trimer_orthogonality-1}
    &+\sum_{\kappa'=0}^{\kappa+k}
    \bra{\phi_\kappa, \varphi_k}\left(\U+\U^2\right)\ket{\phi_{\kappa'}, 
      \varphi_{k'}}
    \braket{\phi_{\kappa'}, \varphi_{k'}}{\psi^\rel_1}
    \Bigg].
\end{align}
Here we have used Eq.~(\ref{eq:energy_conservation}) to eliminate the 
summation over $k'$ in favour of the relationship $k'=\kappa+k-\kappa'$. 
Inserting the {\em ansatz} of Eq.~(\ref{eq:faddeev_component_general_form}) 
to eliminate the Faddeev components $\ket{\psi_\rel}$ and 
$\ket{\widetilde{\psi}_\rel}$ in Eq.~(\ref{eq:trimer_orthogonality-1}) in 
favour of $\ket{f}$ and $\ket{\widetilde{f}}$, respectively, then leads to 
the orthogonality relation:
\begin{align}
  \nonumber
  \braket{\widetilde{\psi}_\rel}{\psi_\rel}=&3\sum_{\kappa,k=0}^\infty 
  \frac{\widetilde{f}_\kappa g_k}
       {(\wErel-E_\kappa-E_k)(\Erel-E_\kappa-E_k)}\\
  &\times \left[
    f_\kappa g_k+\sum_{\kappa'=0}^{\kappa+k}
    \bra{\phi_\kappa, \varphi_k}\left(\U+\U^2\right)\ket{\phi_{\kappa'}, 
      \varphi_{k'}}
    f_{\kappa'} g_{k'}
  \right].
  \label{eq:trimer_orthogonality-2}
\end{align}
In the special case of $\ket{\widetilde{\psi}_\rel}=\ket{\psi_\rel}$ 
Eq.~(\ref{eq:trimer_orthogonality-2}) also yields the normalisation 
constant of the three-body energy state $\ket{\psi_\rel}$.

\subsubsection{Hyper-radial probability density}

A symmetrised zero angular momentum three-body wave function
$\psi_\rel(\vec\rho,\vec r)=\braket{\vec\rho,\vec r}{\psi_\rel}$
depends only on three variables among the six Jacobi coordinates
\cite{sitenko}. These variables may be chosen, for example, as the 
radii $\rho=|\vec \rho|$, $r=|\vec r|$, and the angle between 
$\vec{\rho}$ and $\vec{r}$ \cite{sitenko}. A common way to 
visualise the spatial extent of three-body states by a one dimensional 
function involves the transformation of the Jacobi coordinates $\rho$ and 
$r$ to hyper-spherical coordinates. Among these coordinates the hyper-radius 
and the hyper-angle are given by
\begin{align}
  \label{eq:hyperspherical_coords}
  R(\rho,r)=\sqrt{\frac{m}{\mu_R}}\sqrt{\frac23 \rho^2+\frac12 r^2}
\end{align}
and $\tan\Phi(\rho,r)=\sqrt{\frac43}\,\frac\rho r$, respectively. 
Apart from $R$ all hyper-spherical coordinates are angular variables.
The ``mass'' parameter $\mu_R$ in Eq.~(\ref{eq:hyperspherical_coords})
ensures that $R$ has the unit of a length. In 
Fig.~\ref{fig:trimer_PR_300kHz_158.1G} we have chosen it to be $\mu_R=m$. 
The hyper-radial probability density associated with the three-body 
state $\ket{\psi_\rel}$ is determined, in terms of the projection operator 
\begin{equation}
  \label{eq:projector_PR}
  P_R=\int d^3\rho\  d^3r\ \ket{\vec \rho,\vec r}\ 
  \delta\left(R-R(\rho,r)\right)\ \bra{\vec \rho,\vec r},
\end{equation}
by the formula:
\begin{equation}
  \label{eq:hyperradial_PR}
  P(R)=\bra{\psi_\rel}P_R\ket{\psi_\rel}.
\end{equation}
For a normalised three-body state $P(R)$ satisfies the normalisation 
condition $\int_0^\infty d R\ P(R)=1$ and provides a measure of the  
spatial extent of its wave function.

As the hyper-radius of Eq.~(\ref{eq:hyperspherical_coords}) does not depend on
the angular variables and is also invariant with respect to $\U$, the 
probability density $P(R)$ can be calculated, similarly to 
Eq.~(\ref{eq:trimer_orthogonality-1}), just in terms of the spherically 
symmetric oscillator states $\ket{\phi_{\kappa}}$ and $\ket{\varphi_{k}}$.
This yields
\begin{align}
  \nonumber
  P(R)=&3\sum_{\kappa,k=0}^\infty 
  \bra{\psi^\rel_1}P_R\ket{\phi_\kappa, \varphi_k}
  \Bigg[
    \braket{\phi_\kappa, \varphi_k}{\psi^\rel_1}\\
    \label{eq:hyperradial_PR-1}
    &+\sum_{\kappa'=0}^{\kappa+k}
    \bra{\phi_\kappa, \varphi_k}\left(\U+\U^2\right)
    \ket{\phi_{\kappa'}, \varphi_{k'}}
    \braket{\phi_{\kappa'}, \varphi_{k'}}{\psi^\rel_1}
    \Bigg],
\end{align}
where $k'$ is given by the relationship $k'=\kappa+k-\kappa'$. To determine 
the matrix element
\begin{align}
  \nonumber
  \bra{\psi^\rel_1}P_R\ket{\phi_\kappa, \varphi_k}=&
  \int d^3\rho\  d^3r\ 
  \left[\psi^\rel_1(\vec{\rho},\vec{r})\right]^*\\
  &\times
  \delta\left(R-R(\rho,r)\right) 
  \phi_\kappa(\vec{\rho})\varphi_k(\vec{r}),
  \label{eq:matrix_element_P_R}
\end{align}
we first represent the delta function in Eq.~(\ref{eq:projector_PR}) by the 
formula:
\begin{align}
  \delta(R-R(\rho,r))=\frac{\beta_R^{1/2}}{\beta_r r}\ 
  \delta\left(\beta_r^{-1/2}
      \left(\beta_R R^2-\beta_\rho \rho^2\right)^{1/2} -r\right).
\end{align}
Here we have introduced the parameter $\beta_R=\mu_R\omega_\mathrm{ho}/\hbar$. 
We then analytically perform the five integrations over $r$ and over the 
solid angles associated with $\vec\rho$ and $\vec r$ and substitute in
the remaining integral the variable $\rho$ in accordance with
$\beta_\rho \rho^2=\beta_R R^2w^2$. To represent the matrix element in
Eq.~(\ref{eq:matrix_element_P_R}) in a concise form, we introduce the function
\begin{align}
  \nonumber
  \widehat{\psi}^\rel_1(\beta_R R^2; w^2)=&\sum_{\kappa',k'=0}^\infty
  \sqrt{\frac{\Gamma(\kappa'+1)}{\Gamma(\kappa'+\frac32)}
    \frac{\Gamma(k'+1)}{\Gamma(k'+\frac32)}}
  \frac{f_{\kappa'} g_{k'}}{\Erel-E_{\kappa'}-E_{k'}}\\
  &\times
  \LagP{\kappa'}(\beta_R R^2 w^2)
  \LagP{k'}(\beta_R R^2(1-w^2)),
\end{align}
which can be evaluated numerically. Given the explicit form of the 
harmonic oscillator wave functions in
Eq.~(\ref{eq:wavefunction_HO_one_particle_generic}), the matrix element 
of Eq.~(\ref{eq:matrix_element_P_R}) can then be obtained from the formula:
\begin{align}
  \nonumber
  \bra{\psi^\rel_1}P_R\ket{\phi_\kappa, \varphi_k}=& 
  4\beta_R^{1/2}
  (\beta_R R^2)^{5/2} \exp(-\beta_R R^2)\\
  \nonumber
  &\times\sqrt{\frac{\Gamma(\kappa+1)}{\Gamma(\kappa+\frac32)}
    \frac{\Gamma(k+1)}{\Gamma(k+\frac32)}}
  \int_0^1 d w\ \sqrt{1-w^2}\ w^2\\
  \nonumber
  & \times
  \LagP\kappa(\beta_R R^2 w^2)\LagP{k}(\beta_R R^2(1-w^2))\\
  &\times 
  \widehat{\psi}^\rel_1(\beta_R R^2; w^2).
  \label{eq:hyperradial_PR-3}
\end{align}
In this formula the integration over $w$ lends itself to a numerical
evaluation by a Gauss-Chebyshev quadrature rule of the second kind.

\subsection{Numerical implementation}

\begin{figure}[htbp]
  \includegraphics[width=\columnwidth,clip]{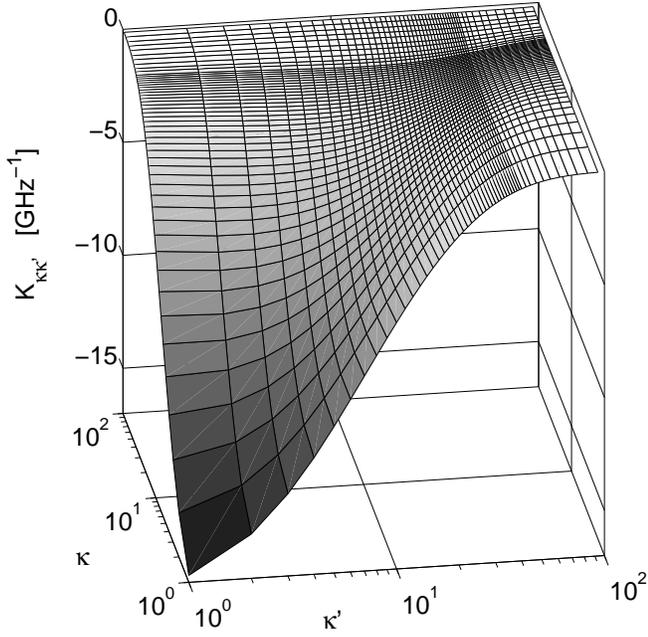}
  \caption{The reduced kernel matrix $\mathcal{K}_{\kappa\kappa'}(\Erel)$ of 
    Eq.~(\ref{eq:faddeev_equation_psi1-coeff-4}), at low vibrational 
    excitations $\kappa,\kappa'<100$, 
    for a $\nu_\mathrm{ho}=300$\,kHz trap and a magnetic field strength of 
    $B=158.1$\,G. 
    The transition from the linear to the connecting exponential mesh at 
    $\kappa=N_\mathrm{e}=40$ is illustrated by the logarithmic scale of the 
    axes associated with the vibrational quantum numbers $\kappa$ and 
    $\kappa'$ (cf.~Eq.~(\ref{eq:kj_sampling})). In the linear part
    ($0\le\kappa,\kappa'<N_\mathrm{e}$) the reduced kernel matrix was 
    evaluated exactly using the results of Appendix \ref{sec:kernel} while the
    approximation of Eq.~(\ref{eq:faddeev_kernel-4}) was used in the
    exponential part. The logarithmic scaling of the axes associated with
    $\kappa$ and $\kappa'$ prevents the row $\kappa=0$ and the column
    $\kappa'=0$ from being displayed.}
  \label{fig:Kkk}
\end{figure}


The analogue of the momentum space Faddeev approach \cite{Gloeckle83} to 
three interacting Bose atoms in a trap involves, contrary to its free-space 
counterpart, discrete Faddeev equations represented by 
Eq.~(\ref{eq:faddeev_equation_psi1-coeff-4}) in the separable potential 
approach. In the limit of low frequency trapping potentials, however, the 
basis set expansion leads to large kernel matrices and the numerical 
determination of the fixed points in 
Eq.~(\ref{eq:faddeev_equation_psi1-coeff-4}) becomes impractical.

\begin{figure}[htbp]
  \includegraphics[width=\columnwidth,clip]{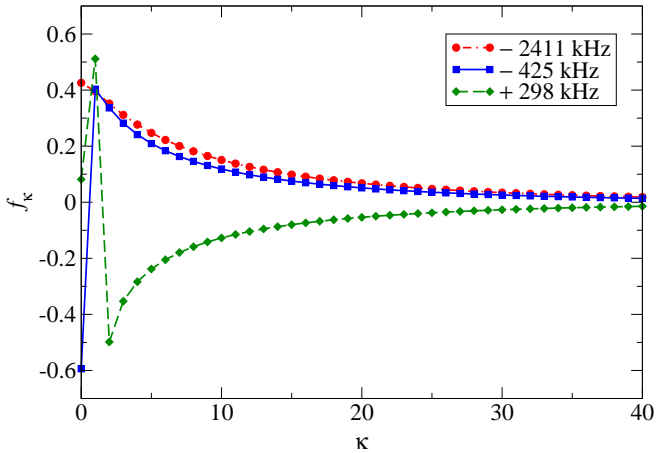}
  \caption{(Color online) Solutions $f_\kappa$ of the matrix equation
    (\ref{eq:faddeev_equation_psi1-coeff-4}), 
    using the reduced kernel matrix shown in
    Fig.~\ref{fig:Kkk}, versus the vibrational quantum number $\kappa$. 
    The legends show the energies $\Etri-\frac62 h\nu_\mathrm{ho}$ of the 
    associated three-body levels of interacting \Rb\ atoms relative to the 
    zero point energy of hypothetically non-interacting atoms in the 
    $\nu_\mathrm{ho}=300$\,kHz trap. 
  }
  \label{fig:fkappa}
\end{figure}

In order to demonstrate the orders of magnitude of the kernel matrix, we 
consider the projection of the form factor $g_k=\braket{\varphi_k}{g}$ onto 
the harmonic oscillator basis. A simple calculation based on 
Eq.~(\ref{eq:formfactor_mom}) shows that this projection is given by the 
formula:
\begin{equation}
  \label{eq:formfactor_HO}
  g_k=\frac{4}{\pi^{1/4}} \frac{\zeta^{3/4}}{(1+\zeta)^{3/2}}
  \left(\frac{1-\zeta}{1+\zeta}\right)^k 
  \sqrt{\frac{\Gamma(k+\frac32)}{\Gamma(k+1)}}.
\end{equation}
Here we have introduced the dimensionless parameter 
$\zeta=\beta_r \sigma^2=\sigma^2 \frac{1}{2}m\omega_\mathrm{ho}/\hbar$.
The form factor thus decays like $g_k\sim e^{-2k\zeta}$ in the limit of large 
vibrational quantum numbers $k$. In the case of $^{85}$Rb atoms the parameter 
$\zeta$ is on the order of $1.7\times 10^{-2}$ for a 
$\nu_\mathrm{ho}=300$\,kHz atom trap, while it gets as small as 
$1.1\times 10^{-5}$ at a trap frequency of $\nu_\mathrm{ho}=200$\,Hz. 
If we estimated the order of magnitude of a numerical cut-off $k_\mathrm{max}$,
for instance, by supposing that the form factor is well represented when its 
value $g_k$ at $k=k_\mathrm{max}$ has decayed to $10^{-3} g_0$ we would 
require only $k_\mathrm{max}\approx200$ basis states in the case of a  
300\,kHz trap. For a 1\,kHz trap, however, this number would increase to 
$k_\mathrm{max}\approx 320000$. At low trap frequencies the kernel matrix 
would, therefore, become too large for a numerical solution of
Eq.~(\ref{eq:faddeev_equation_psi1-coeff-4}) to be practical.

To account for the discrete nature of the basis set expansion on the one hand
and the exponential scaling of $g_k$, in the limit of large $k$, 
on the other hand, we have introduced a 
linear mesh of length $N_\mathrm{e}=40$ covering each of the lowest energy 
states, and a connecting exponential mesh of the same length containing the 
energetically higher states in the following way:
\begin{equation}
  \label{eq:kj_sampling}
  k_j=\left\{
  \begin{array}{lcc}
    j &:& 0 \le j < N_\mathrm{e}\\
    N_\mathrm{e}+\lfloor\ 
    \left[ e^{(j-N_\mathrm{e}) \xi \delta}-1\right]/\xi\ \rfloor
    &:& N_\mathrm{e}\le j<N_\mathrm{t}\\
  \end{array}
  \right..
\end{equation}
Here the symbol $\lfloor x \rfloor$ indicates the largest integer less than 
or equal to $x$, and $\xi\ll1$ and $\delta$ are adjustable parameters. The
parameter $\delta$ can be interpreted as the initial step size at
the transition from the linear to the exponential mesh,
i.e.~$k_{N_\mathrm{e}+1}-k_{N_\mathrm{e}} \approx\delta $, which we 
have chosen as $\delta=2$. The parameter $\xi$ is fixed by the requirement
$k_{N_\mathrm{t}-1}=k_\mathrm{max}$, leading to a transcendental equation for
$\xi$ which can be solved numerically. In our applications the complete 
mesh then consisted of $N_\mathrm{t}=80$ points.
We have generated an equivalent mesh for the sampling of the vibrational 
quantum number $\kappa_j$, resulting in an $80\times80$ reduced kernel matrix 
$\mathcal{K}_{\kappa\kappa'}(\Erel)$, which has been used for the numerical
solution of Eq.~(\ref{eq:faddeev_equation_psi1-coeff-4}). 
Figure (\ref{fig:Kkk}) shows the reduced kernel matrix for a 
$\nu_\mathrm{ho}=300$\,kHz atom trap at a magnetic field strength of
$B=158.1$\,G. The solutions $f_\kappa$ of 
Eq.~(\ref{eq:faddeev_equation_psi1-coeff-4}) for the three lowest energetic 
three-body states obtained with the reduced kernel matrix of 
Fig.~\ref{fig:Kkk} are illustrated in Fig.~\ref{fig:fkappa}.

\subsection{Accuracy of the separable potential approach with respect to 
  three-body energy spectra}
The range of validity of our exact solutions to the three-body Faddeev
equations is limited by the accuracy of the inter-atomic potentials. 
Figure \ref{fig:dimer_energies_300kHz} clearly demonstrates that the 
simultaneous adjustment of the separable potential of Appendix 
\ref{App:Separable_potential} to the scattering length of 
Eq.~(\ref{eq:scattering_length_B_field}) and to the formula 
(\ref{eq:E_2BGF}) for the binding energy of an alkali van der 
Waals molecule \cite{GF_PRA48} accurately describes both the 
measurements of Ref.~\cite{Claussen03} and their {\em ab initio} theoretical 
predictions \cite{Kokkelmans_private04}. This accuracy of the approach 
with respect to the energies of $^{85}$Rb$_2$ Fesh\-bach molecules 
persists over a wide range of magnetic field strengths, extending from the 
position of the zero energy resonance of about 155 G up to 161\,G, which is
far beyond the range of validity of the universal formula 
(\ref{eq:near_resonant_binding_energy}). Our separable potential approach 
also recovers the excited state energy spectra for a pair of $^{85}$Rb atoms 
in a 300\,kHz trap which we have determined using a microscopic binary 
interaction $V(\mathbf{r})$ (see Fig.~\ref{fig:dimer_energies_300kHz}).

To demonstrate the validity of our separable potential approach 
also in its applications to three-body energy spectra, we shall compare its 
predictions to the {\em ab initio} binding energies of $^{4}$He$_3$ provided 
in Ref.~\cite{Barletta01}. As the low static electric dipole polarizability of 
helium and the correspondingly small van der Waals coefficient of 
$C_6=1.461\,$a.u.~\cite{Tang95} do not allow us to accurately recover the 
helium dimer binding energy from Eq.~(\ref{eq:E_2BGF}), we have used 
$E_\mathrm{b}/k_\mathrm{B}=-1.313\,$mK 
($k_\mathrm{B}=1.3806505\times 10^{-23}$ J/K is the Boltzmann constant)
in addition to the scattering length of $a=190.7\,$a.u.~to adjust  
the parameters $A$ and $\sigma$ of the separable potential of Appendix 
\ref{App:Separable_potential}. These particular values of $a$ and 
$E_\mathrm{b}$ correspond to the Tang, Toennies and Yiu (TTY) helium dimer 
potential \cite{Tang95} reported in Ref.~\cite{Barletta01} and determine the 
range parameter of the separable potential to be $\sigma=8.02\,$a.u.. This 
approach predicts the $^4$He$_3$ ground state energy to be 
$E_1^\mathrm{sep}/k_\mathrm{B}=-96.9\,$mK and for the excited Efimov 
state we obtain $E_2^\mathrm{sep}/k_\mathrm{B}=-2.09$ mK. These 
predictions only slightly overestimate the exact energies of 
$E_1^\mathrm{TTY}/k_\mathrm{B}=-126.4\,$mK and
$E_2^\mathrm{TTY}/k_\mathrm{B}=-2.277\,$mK \cite{Barletta01} with 
relative deviations of 23\,\% and 8\,\%, respectively.

To study the uncertainties of our approach to determine the separable 
interaction, we have also performed a different adjustment of $V_\mathrm{sep}$ 
based on the requirement that its effective range 
$r_\mathrm{eff}^\mathrm{sep}=(4\sigma/\sqrt{\pi})[1-\sqrt{\pi}\sigma/(2a)]$
exactly recovers the effective range of 
$r_\mathrm{eff}=13.85\,$a.u.~of the TTY potential in addition to the exact 
scattering length. This determines the range parameter to be 
$\sigma=6.31\,$a.u.. The associated potential $V_\mathrm{sep}$ yields 
$E_\mathrm{b}^\mathrm{sep}/k_\mathrm{B}=-1.285\,$mK for the helium 
dimer, while it predicts the trimer energies to be 
$E_1^\mathrm{sep}/k_\mathrm{B}=-142.6\,$mK and 
$E_2^\mathrm{sep}/k_\mathrm{B}=-2.36\,$mK with relative deviations 
from the {\em ab initio} results \cite{Barletta01} of 12\,\% and 4\,\%, 
respectively. Both adjustments of the separable potential thus lead to 
similar degrees of accuracy, which are comparable to the accuracy of the 
adiabatic hyper-spherical approach of Ref.~\cite{Esry96}. 

The small discrepancies between the predictions obtained from the different 
separable potentials and the {\em ab initio} calculations of 
Ref.~\cite{Barletta01} indicate a remnant sensitivity of the trimer binding 
energies to properties of the microscopic binary interactions beyond 
those accounted for by the second order effective range expansion of 
the two-body scattering phase shift. 
Such corrections can, in principle, be exactly included in our calculations 
by taking advantage of the universal properties of low energy three-body 
spectra discussed, e.g., in Ref.~\cite{Braaten03}. This 
presupposes, however, that one of the trimer binding energies is known either 
from experiment or from {\em ab initio} calculations. For instance, adjusting 
$V_\mathrm{sep}$ in such a way that it exactly recovers the trimer ground 
state energy of Ref.~\cite{Barletta01} in addition to the binary scattering 
length yields $\sigma=6.79\,$a.u., which recovers the exact energy of the 
excited Efimov state reported in Ref.~\cite{Barletta01} to a relative accuracy 
of 0.5\,\%. The studies of Ref.~\cite{Gdanitz01} suggest, however, that the 
relatively small deviations between the {\em ab initio} and the different 
approximations employed in the separable potential approach may be physically 
insignificant due to the remaining uncertainties of even the comparatively 
well known helium dimer interaction potential.


\begin{thebibliography}{3}
\bibitem{Borromean}
  This terminology refers to the emblem of the Italian aristocratic family 
  Borromeo, which consists of three rings interlinked in such a way that 
  the remaining rings fall apart as soon as any one of the rings is removed.
\end{thebibliography}
\end{document}